\documentclass[9pt,twocolumn,twoside]{osajnl}

\journal{ao} 

\setboolean{shortarticle}{false}

\usepackage{subfigure}
\usepackage{lineno}

\DeclareMathOperator*{\argmin}{arg\,min}

\title{Single-shot blind deconvolution with coded aperture}

\author[1]{Hideyuki Muneta}
\author[2,*]{Ryoichi Horisaki}
\author[3]{Yohei Nishizaki}
\author[2]{Makoto Naruse}
\author[1]{Jun Tanida}

\affil[1]{Department of Information and Physical Sciences, Graduate School of Information Science and Technology, Osaka University, 1-5 Yamadaoka, Suita, Osaka 565-0871, Japan}
\affil[2]{Department of Information Physics and Computing, Graduate School of Information Science and Technology, The University of Tokyo, 7-3-1 Hongo, Bunkyo-ku, Tokyo 113-8656, Japan}
\affil[3]{Systems and Control Laboratory, Environmental Technology Research Division, Osaka Research Institute of Industrial Science and Technology, 1-6-50, Morinomiya, Joto-ku, Osaka 536-8553, Japan}

\affil[*]{Corresponding author: horisaki@g.ecc.u-tokyo.ac.jp}

\begin{abstract}
In this paper, we present a method for single-shot blind deconvolution incorporating a coded aperture~(CA).
In this method, we utilize the CA, inserted on the pupil plane, as support constraints in blind deconvolution.
Not only an object but also a point spread function of turbulence are estimated from a single captured image by a reconstruction algorithm with the CA support.
The proposed method is demonstrated by a simulation and an experiment in which point sources are recovered under severe turbulence.
\end{abstract}

\setboolean{displaycopyright}{false}

\begin{document}

\maketitle

\section{Introduction}

Blind deconvolution is a category of deconvolution methods and has been studied for optical imaging through turbulence for astronomy, biomedicine, security, and so on~\cite{Ayers1988, Schulz1993, Kundur1996, Chaudhuri2014}.
In non-blind deconvolution, an object image is retrieved from a known point spread function~(PSF) and the captured image.
On the other hand, blind deconvolution estimates both an object image and a PSF from a single or multiple captured images.
Blind deconvolution is in high demand in a wide range of applications, as mentioned above, compared to non-blind deconvolution because PSFs may be unknown in many practical situations.
However, blind deconvolution is an ill-posed problem, especially in the case of single-shot observations, and it is difficult to solve stably.
Therefore, conventionally, blind deconvolution requires various assumptions to be valid, such as a limited size of PSFs~(i.e. weak aberrations) and certain prior knowledge about the imaging conditions.

Coded apertures~(CAs) are typically used in computational imaging and have been applied to both non-blind and blind deconvolutions~\cite{ables1968, Dicke1968, Fenimore1978}.
In the case of non-blind deconvolution, CAs have been employed for improving the condition of inverse defocus problems in imaging systems~\cite{Veeraraghavan2007,Levin2007,Zhou2011,Asif2017, Horisaki2020}.
Blind deconvolution based on CAs has also been reported with multi-shot modalities for imaging through turbulence, where a single CA or multiple CAs are assumed.
Aperture masking interferometry in astronomy is a single-CA-based approach for compensating for atmospheric turbulence~\cite{Labeyrie1970, Baldwin1986, Haniff1987, Haniff1992, Peng2017, Sallum2017}.
This method observes a large number of short-exposure images of a stationary object through different scattering processes~(i.e.~dynamically changing turbulence) by using a single CA for estimating the object's spectrum or the object itself.
In aperture masking interferometry, the pinholes on the CA must be aligned as non-redundant configurations.
This means that the pinholes on the CA are very sparsely arranged;~hence the light efficiency is low in this method.
The multiple-CAs-based approach also has been used for blind deconvolution, where the object is observed through a stationary scattering process with different CAs sequentially~\cite{Horstmeyer2014, Chung2019}.
As mentioned so far, a serious issue of these blind deconvolution methods in the literature is the limited capability in imaging dynamically changing objects due to the multi-shot CA modalities.

In this paper, we present a method for single-shot blind deconvolution using a CA with high light efficiency.
We consider the CA as a support of the pupil function during the blind deconvolution process.
By using the CA support, the number of estimated variables on the pupil plane is reduced by partially blocking the light from the object, and the condition of the inverse problem is improved. 
This method may allow us to solve the single-shot blind deconvolution problem stably by using not only computations but also optical hardware.
Our study will contribute to imaging through turbulence in various applications by enhancing imaging speeds, tolerating strong turbulence, and simplifying optical hardware.

\section{Method}

\begin{figure}
\begin{center}
\includegraphics[scale=0.6]{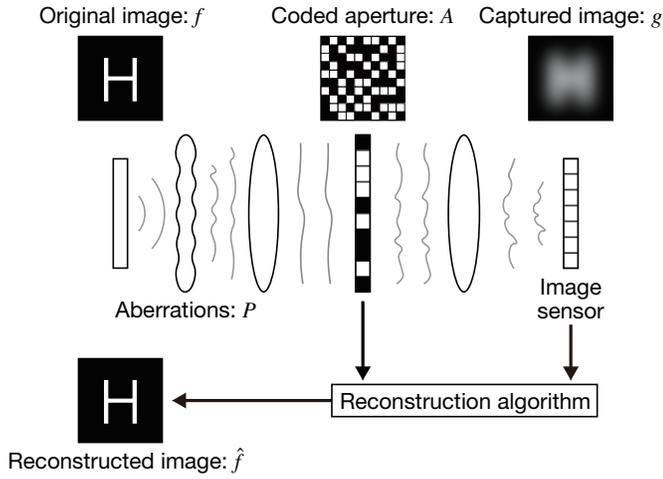}
\end{center}
\caption{Schematic diagram of single-shot blind deconvolution with a coded aperture~(CA).}
\label{methodsystem}
\end{figure}

A diagram of the proposed method is shown in Fig.~\ref{methodsystem}.
In our method, an object~$f$ illuminated with spatially incoherent light is captured as a single intensity image~$g$ through turbulence and the CA.
The CA is located on the pupil plane in the imaging optics and is described as a binary random pattern~$A$.
By assuming shift-invariant aberration of the turbulence, the forward model is written with a convolution process as
\begin{align}
\begin{split}
&g=f\otimes h_{\rm i},\\
&h_{\rm i}=|\mathcal{F}^{-1}[P\times A]|^2,
\label{eq_forward}
\end{split}
\end{align}
where $h_{\rm i}$ is the point spread function~(PSF) of incoherent light through the turbulence and the imaging optics, and $P$ is the transfer function of the aberrations by the turbulence under coherent light, which we call the aberration pupil function.
These variables are two-dimensional.
Lowercase variables are ones in the spatial domain, and uppercase variables are ones in the Fourier domain, respectively.
Here, $\otimes$ denotes the convolution operator, $\times$ denotes the element-wise product operator, and $\mathcal{F}^{-1}$ is the inverse Fourier transform, respectively.

The inverse problem of the single-shot blind deconvolution with the CA in Eq.~(\ref{eq_forward}) is written as
\begin{align}
\widehat{f}=\argmin_{\widehat{f},\widehat{P}(\{u|A(u)=1\})}\|g-\widehat{f}\otimes|\mathcal{F}^{-1}[\widehat{P}\times A]|^2 \|_2^2,
\label{eq_inv}
\end{align}
where $u$ is a set of frequency coordinates, $\|\bullet\|_2$ is the $\ell_2$-norm, and the accent symbol on $\widehat{\quad}$ denotes estimated variables.
Instead of estimating the PSF~$h$ in the conventional blind deconvolution, we retrieve the aberration pupil function~$P$ on the binary CA support~$A$ simultaneously with the reconstruction of the object~$f$.
In this case, the estimated variables on the aberration pupil function~$P$ are only at the frequency coordinates where the light passes through the CA~($A(u)=1$), and variables at the frequency coordinates where the light is blocked by the CA~($A(u)=0$) do not need to be estimated.
Therefore, we can limit the number of estimated variables by using the support~$A$ even when the PSF~$h$ is large due to strong aberrations.

\begin{figure}
\begin{center}
\includegraphics[scale=0.6]{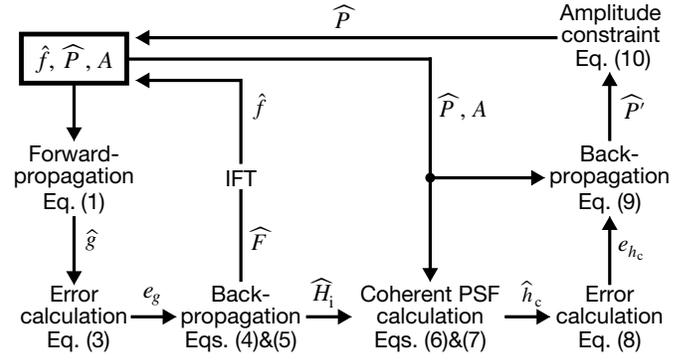}
\end{center}
\caption{Flow of the reconstruction algorithm.
$e_g$:~the error of the captured image.
$\widehat{F}$:~the estimated Fourier spectrum of $\widehat{f}$.
$\widehat{H_{\rm i}}$:~the estimated incoherent transfer function.
$\widehat{h_{\rm c}}$:~the estimated coherent PSF.
$e_{h_{\rm c}}$:~the error of the coherent PSF.
IFT:~the inverse Fourier transform.
}
\label{optalgorithm}
\end{figure}

We solve the inverse problem in Eq.~(\ref{eq_inv}) based on the ptychographic iterative engine~(PIE), which has been used for coherent diffractive imaging~\cite{Maiden2012,Horisaki2019}.
The flow of the reconstruction is shown in Fig.~\ref{optalgorithm}.
We initially set the estimated original image~$\widehat{f}$ and the estimated aberration pupil function~$\widehat{P}$, and calculate the simulated captured image~$\widehat{g}$ by using the forward model of Eq.~(\ref{eq_forward}).
Then, the error~$e_g$ between the real and simulated captured images is computed as
\begin{align}
e_g = g-\widehat{g}.
\label{eq_loss}
\end{align}
Based on the PIE, this error separately back-propagates to the Fourier spectrum~$\widehat{F}$ of the estimated object image and the estimated transfer function~$\widehat{H_{\rm i}}$ of incoherent light, which is the Fourier transform of $\widehat{h_{\rm i}}$, as
\begin{align}
&\widehat{F}=\widehat{F}+\frac{\widehat{H_{\rm i}}^* \times \mathcal{F}[e_g]}{(1-{\alpha}_F) |\widehat{H_{\rm i}}|^2 + {\alpha}_F {|\widehat{H_{\rm i}}|^2_{\rm max}}},
\label{eq_PIE_F}\\
&\widehat{H_{\rm i}}=\widehat{H_{\rm i}}+\frac{\widehat{F}^* \times \mathcal{F}[e_g]}{(1-{\alpha}_{H_{\rm i}}) |\widehat{F}|^2 + {\alpha}_{H_{\rm i}} {|\widehat{F}|^2_{\rm max}}},
\label{eq_PIE_H_i}
\end{align}
where $\mathcal{F}$ is the Fourier transform, the superscript of $^*$ denotes the complex conjugation operator, and the subscript of $\rm max$ is the maximum value on an image.
The parameters~$\alpha_F$ and $\alpha_{H_{\rm i}}$ are constants for the PIE-based feedback to $\widehat{F}$ and $\widehat{H_{\rm i}}$, respectively.
The estimated object image~$\widehat{f}$ is the inverse Fourier transform of $\widehat{F}$ in Eq.~(\ref{eq_PIE_F}).

Next, we estimate the aberration pupil function~$\widehat{P}$, as shown in the right half-plane of Fig.~\ref{optalgorithm}.
The tentative PSF~$\widehat{h_{\rm c}}'$ of coherent light is described with $\widehat{P}$ and $A$ as
\begin{align}
\widehat{h_{\rm c}}'=\mathcal{F}^{-1}[\widehat{P}\times A].
\label{eq_h_c_p}
\end{align}
Its rectified PSF~$\widehat{h_{\rm c}}$ is calculated with an amplitude constraint with $\widehat{H_{\rm i}}$ in Eq.~(\ref{eq_PIE_H_i}) as
\begin{align}
\widehat{h_{\rm c}}=\sqrt{\mathcal{F}^{-1}[\widehat{H_{\rm i}}]}{\rm exp}(i \times {\rm arg}(\widehat{h_{\rm c}}')),
\label{eq_phase}
\end{align}
where $i$ is the imaginary number, and ${\rm arg}(\bullet)$ is the argument of complex variables.
The error~$e_{h_{\rm c}}$ between $\widehat{h_{\rm c}}$ and $\widehat{h_{\rm c}}'$ is written as
\begin{align}
e_{h_{\rm c}} = \widehat{h_{\rm c}}-\widehat{h_{\rm c}}',
\label{eq_lossh}
\end{align}
and it back-propagates to the tentative aberration pupil~function~$\widehat{P}'$ as 
\begin{align}
\widehat{P}'=\widehat{P}+\frac{A^* \times \mathcal{F}[e_{h_{\rm c}}]}{(1-{\alpha}_P) |A|^2 + {\alpha}_P {|A|^2_{\rm max}}},
\label{eq_PIEh}
\end{align}
where the parameter~$\alpha_P$ is a constant for the PIE-based feedback to $\widehat{P}'$.
The rectified aberration pupil function~$\widehat{P}$ of $\widehat{P}'$ is calculated with an amplitude constraint by assuming turbulence with no light absorption~(or spatially invariant light absorption) as
\begin{align}
\widehat{P}={\rm exp}(i \times {\rm arg}(\widehat{P}')).
\label{eq_ampconst}
\end{align}
This process is iterated until $\|e_g\|_2^2$ converges.

\section{Demonstration}

\begin{figure}[t!]
\begin{center}
		\subfigure[]{\label{sim_or}\includegraphics[height=2.8cm]{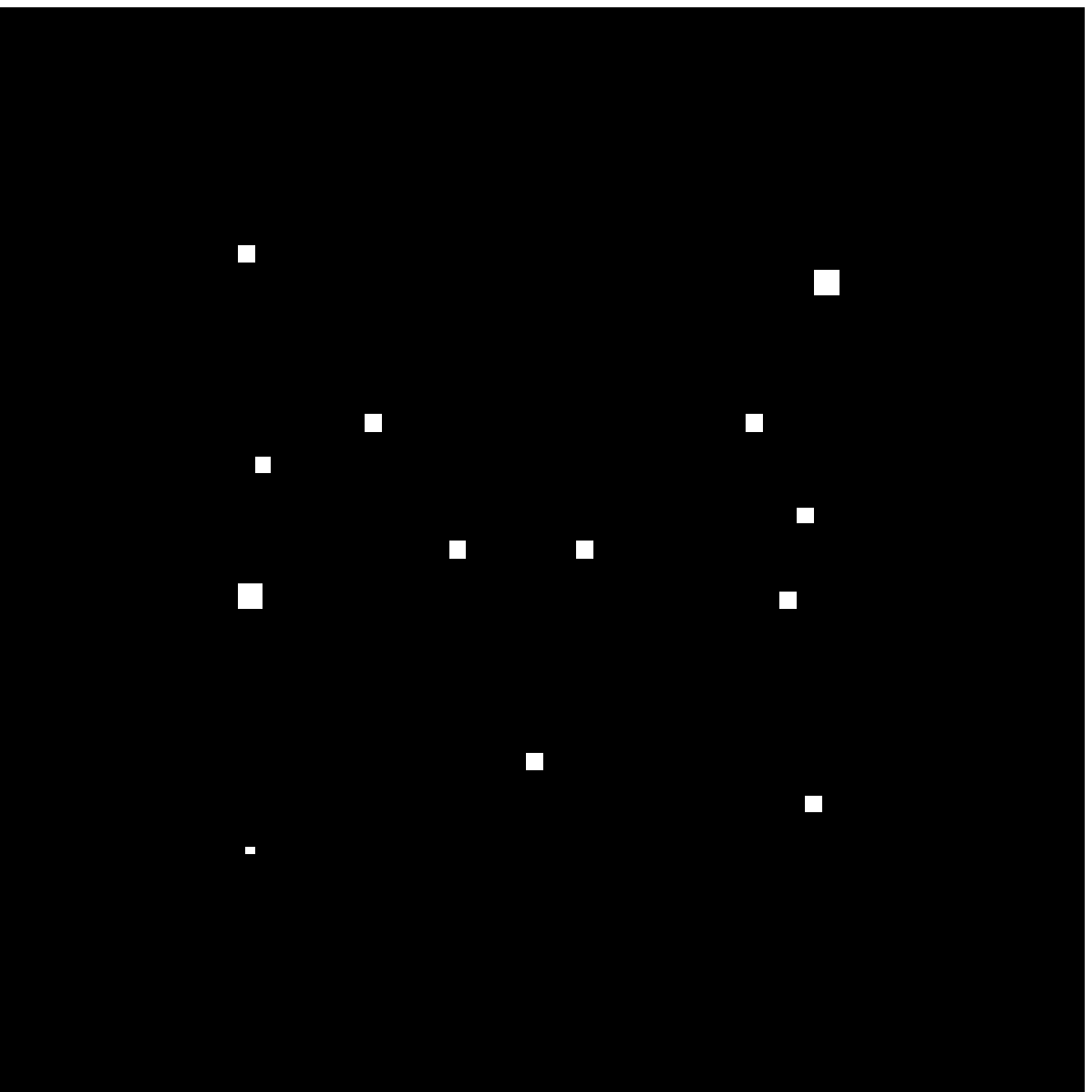}}
		\hspace{0.5cm}
		\subfigure[]{\label{sim_CA}\includegraphics[height=2.8cm]{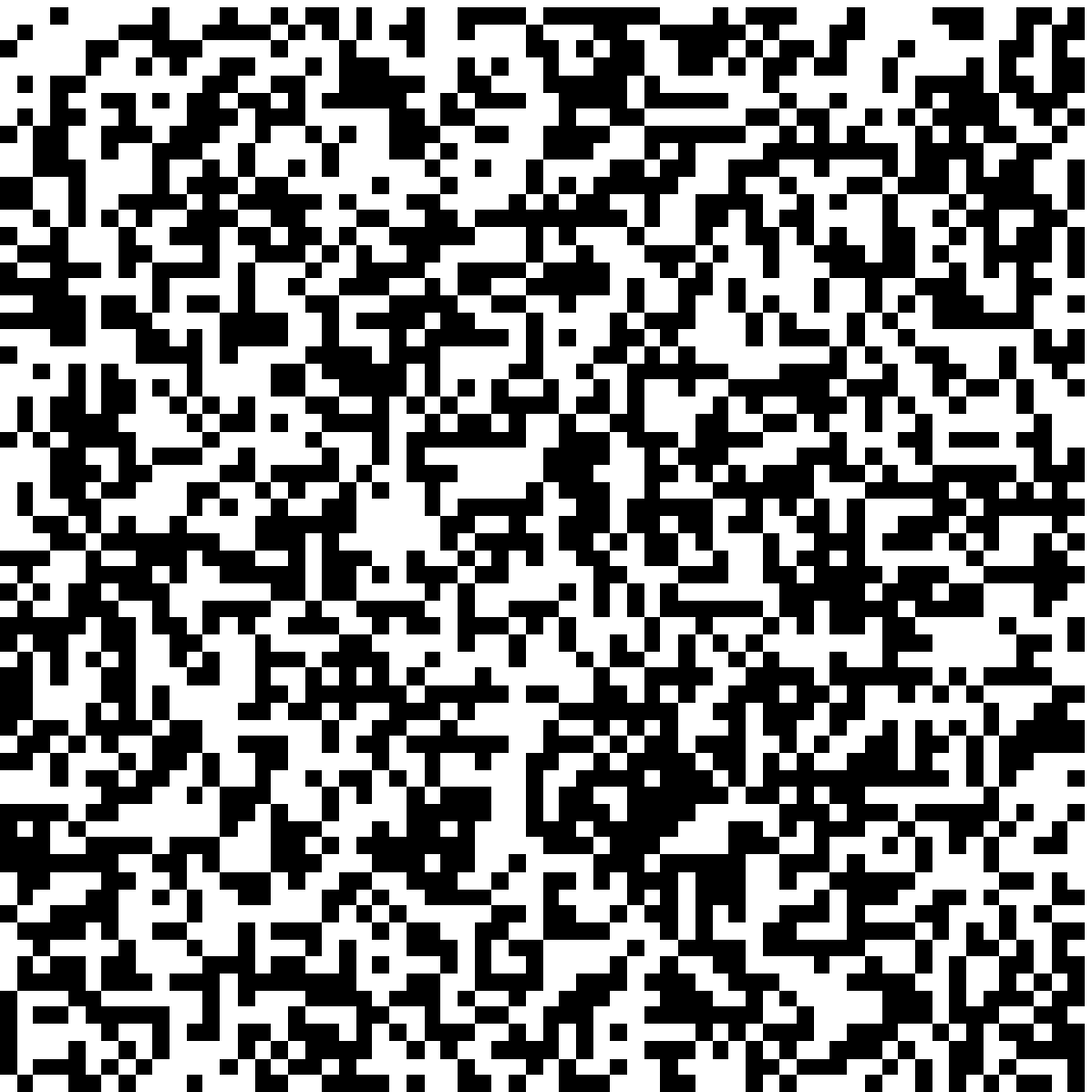}}\\
		\subfigure[]{\label{sim_Z8_map}\includegraphics[height=2.8cm]{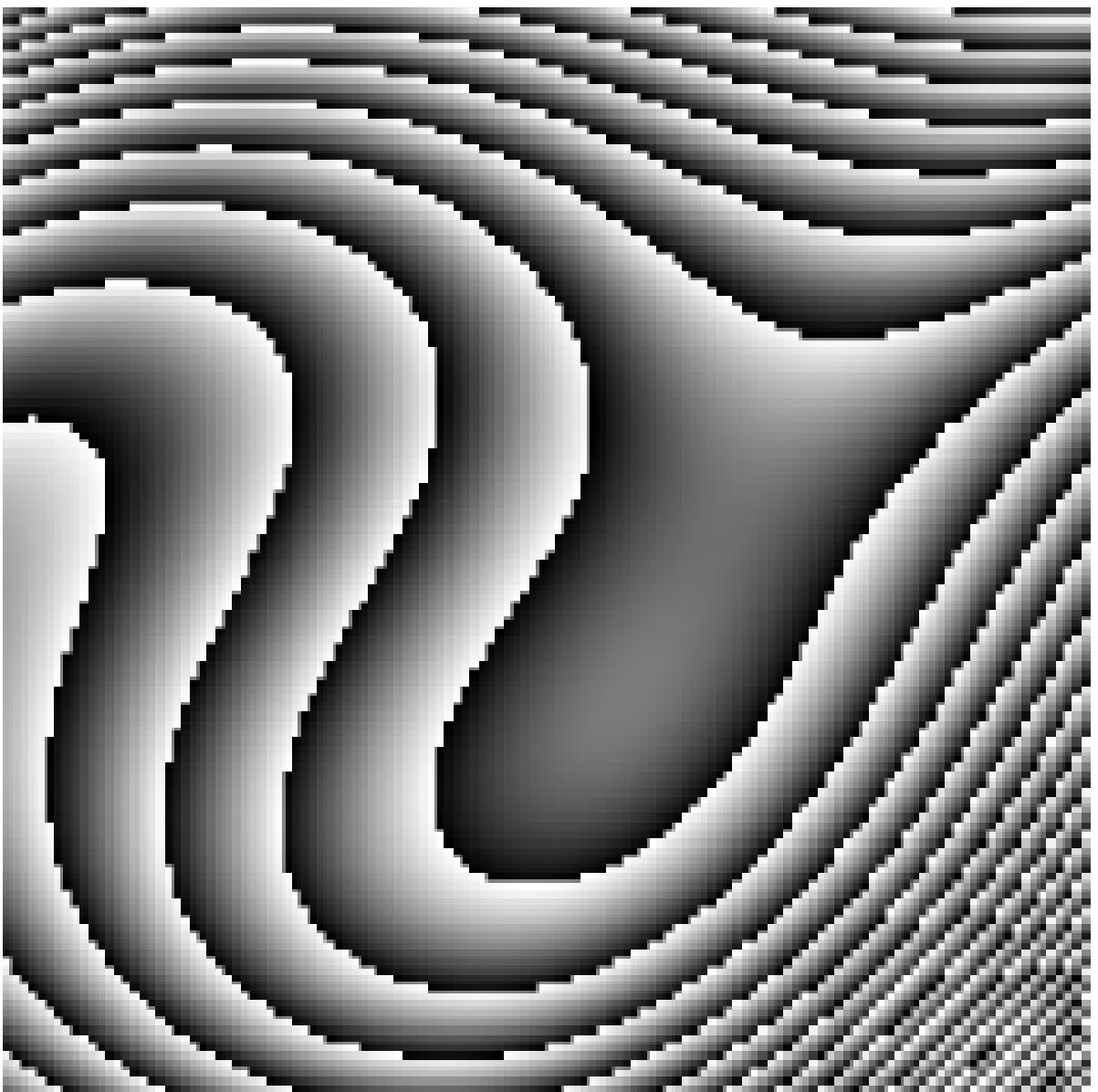}}
		\subfigure[]{\label{sim_Z16_map}\includegraphics[height=2.8cm]{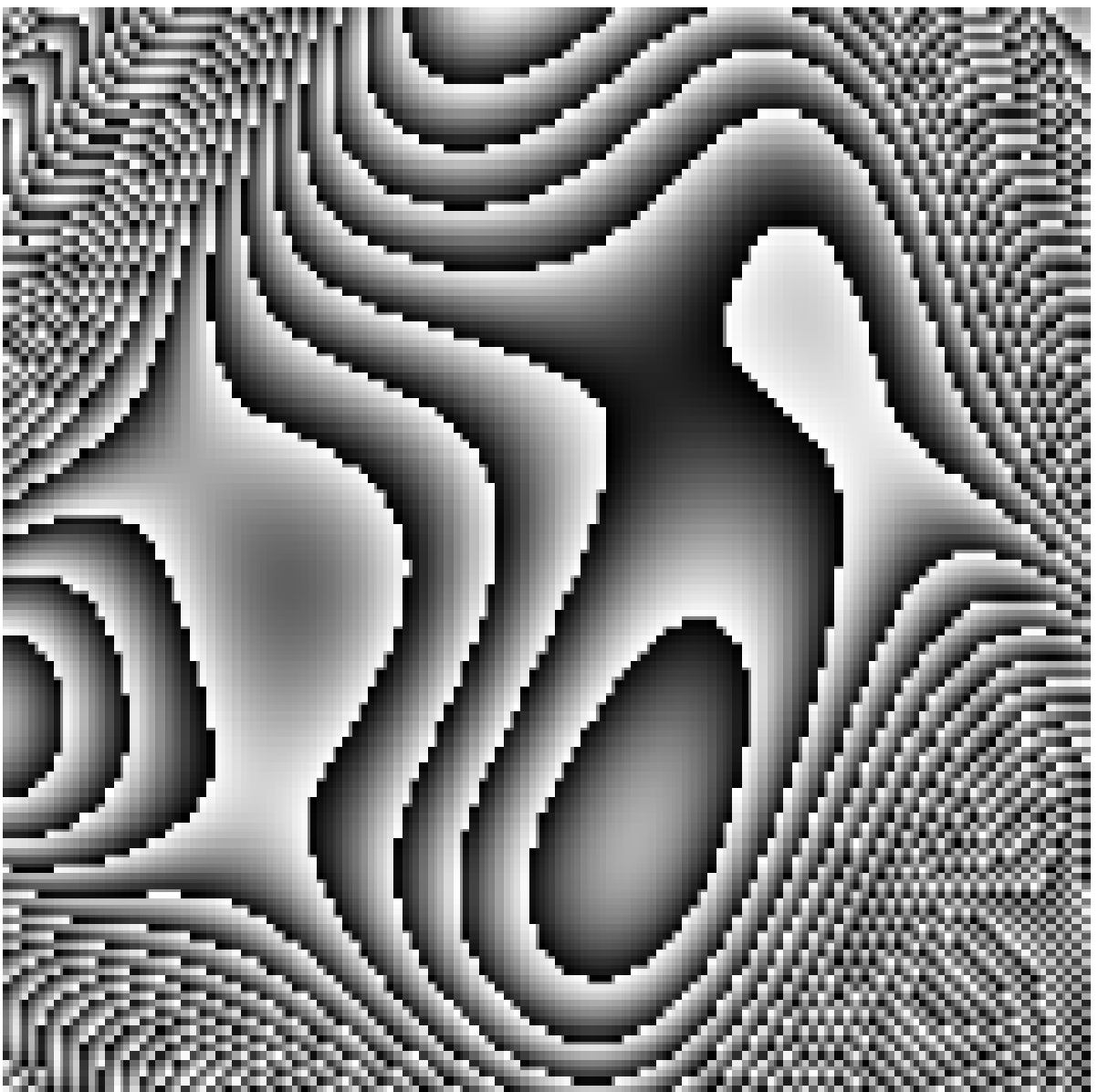}}
		\subfigure[]{\label{sim_Z24_map}\includegraphics[height=2.8cm]{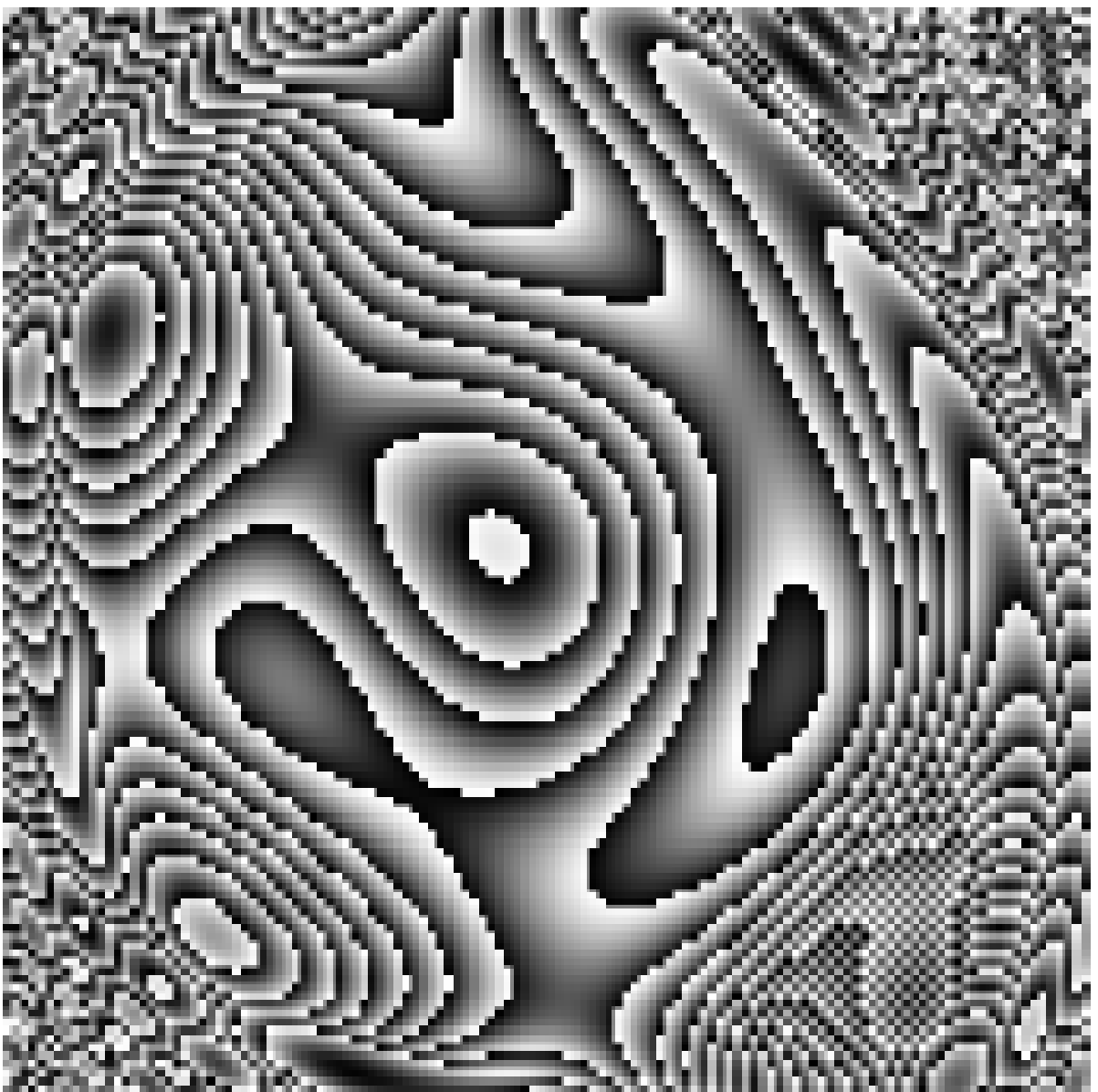}}\\
		\subfigure[]{\label{sim_Z8_cap}\includegraphics[height=2.8cm]{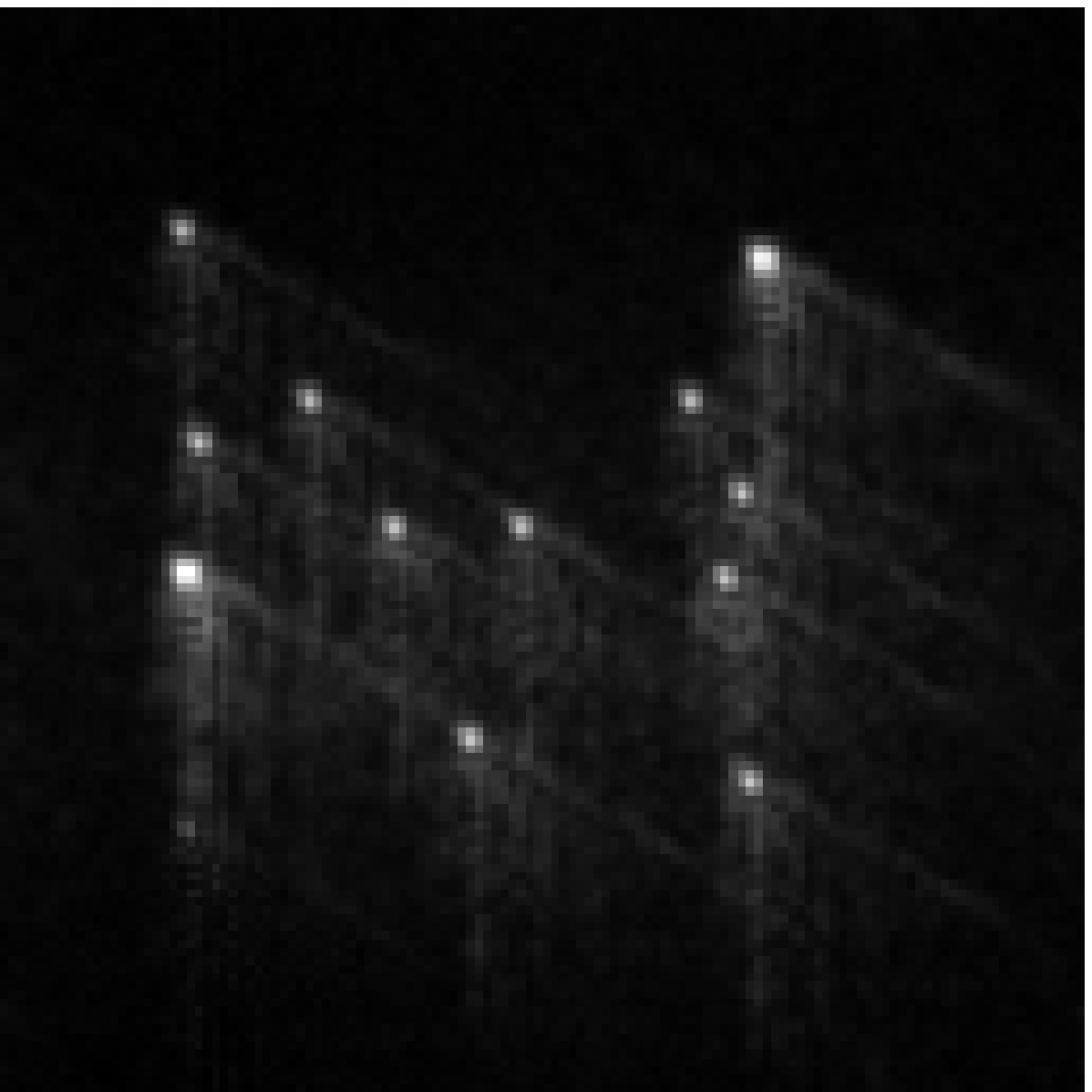}}
		\subfigure[]{\label{sim_Z16_cap}\includegraphics[height=2.8cm]{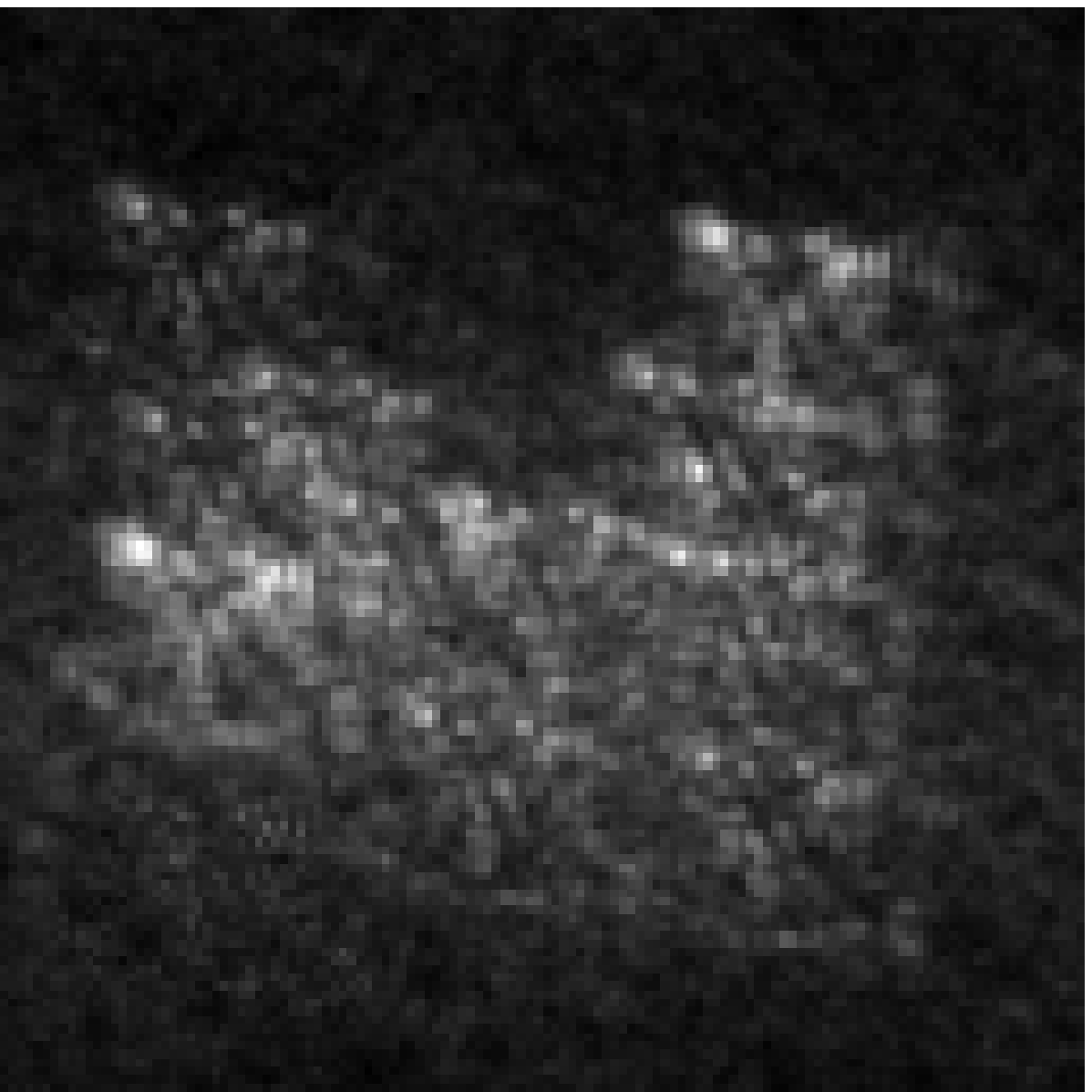}}
		\subfigure[]{\label{sim_Z24_cap}\includegraphics[height=2.8cm]{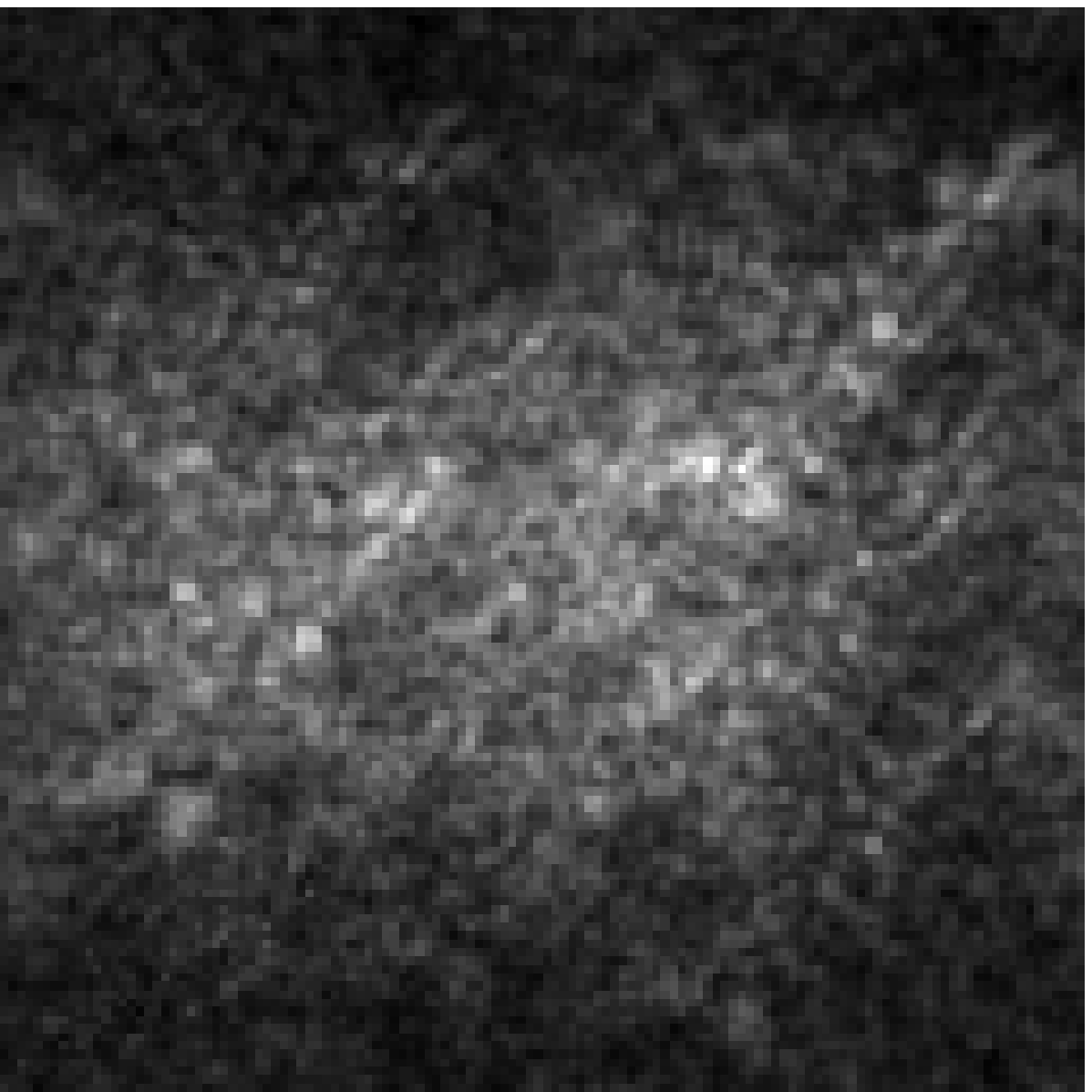}}\\
		\subfigure[]{\label{sim_Z8_rec}\includegraphics[height=2.8cm]{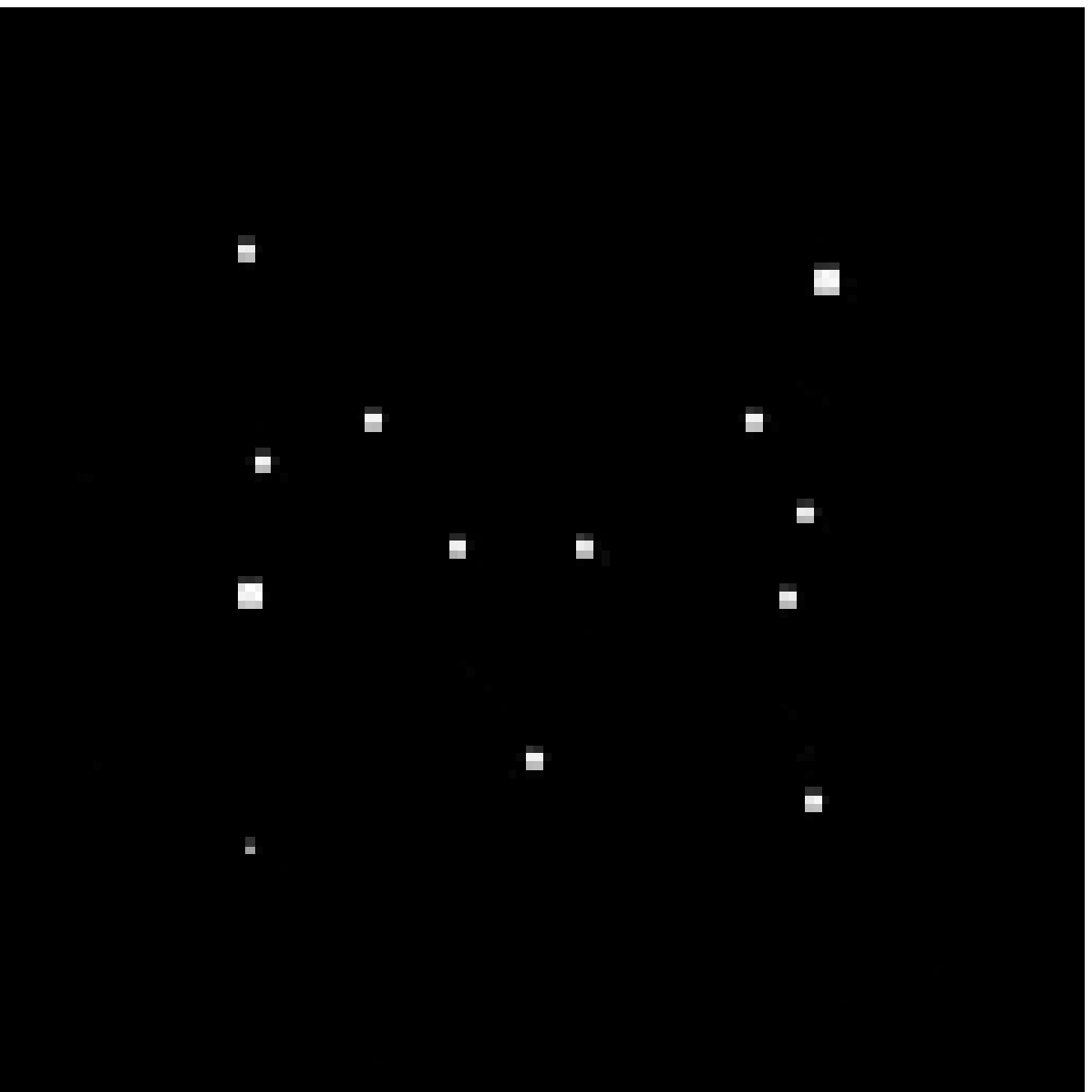}}
		\subfigure[]{\label{sim_Z16_rec}\includegraphics[height=2.8cm]{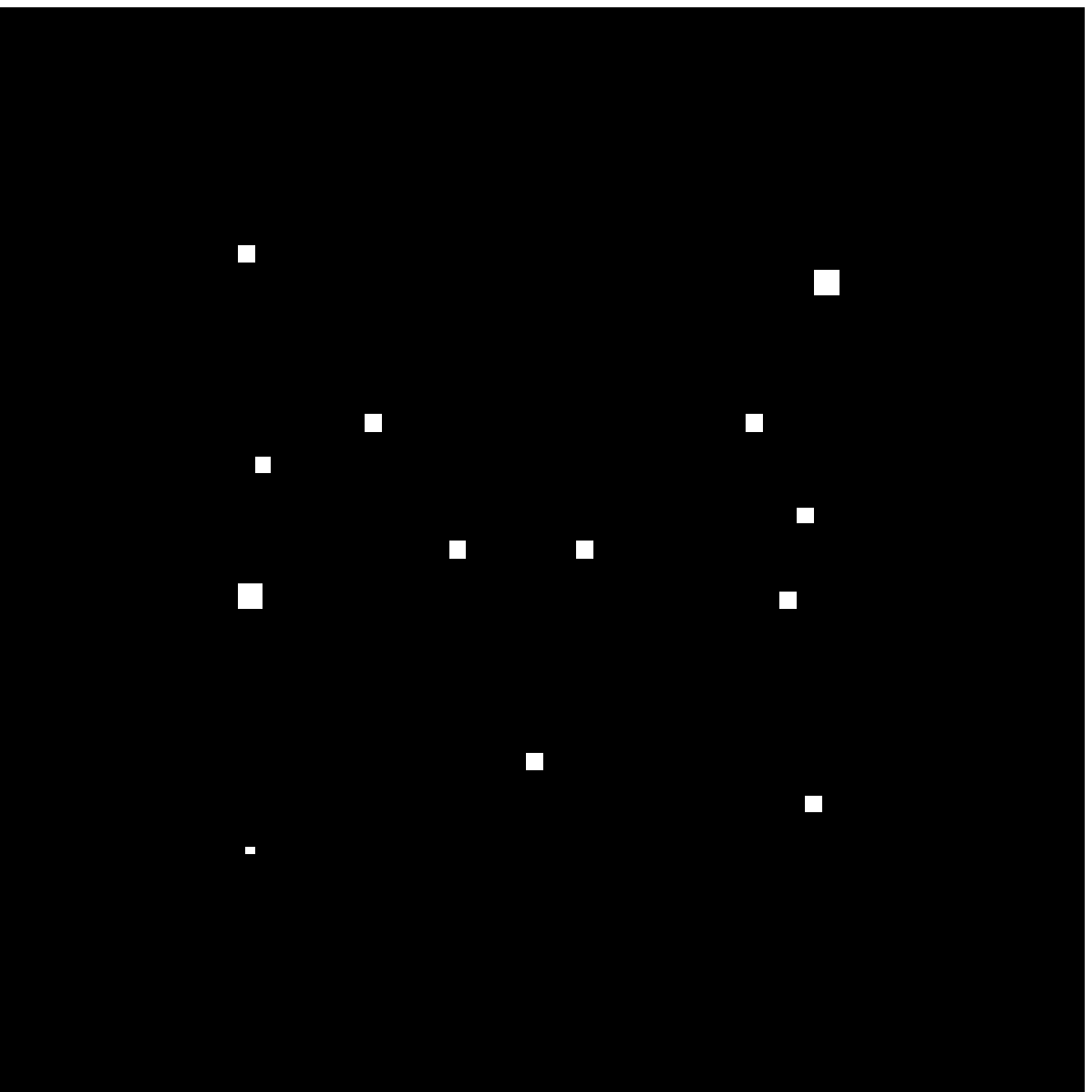}}
		\subfigure[]{\label{sim_Z24_rec}\includegraphics[height=2.8cm]{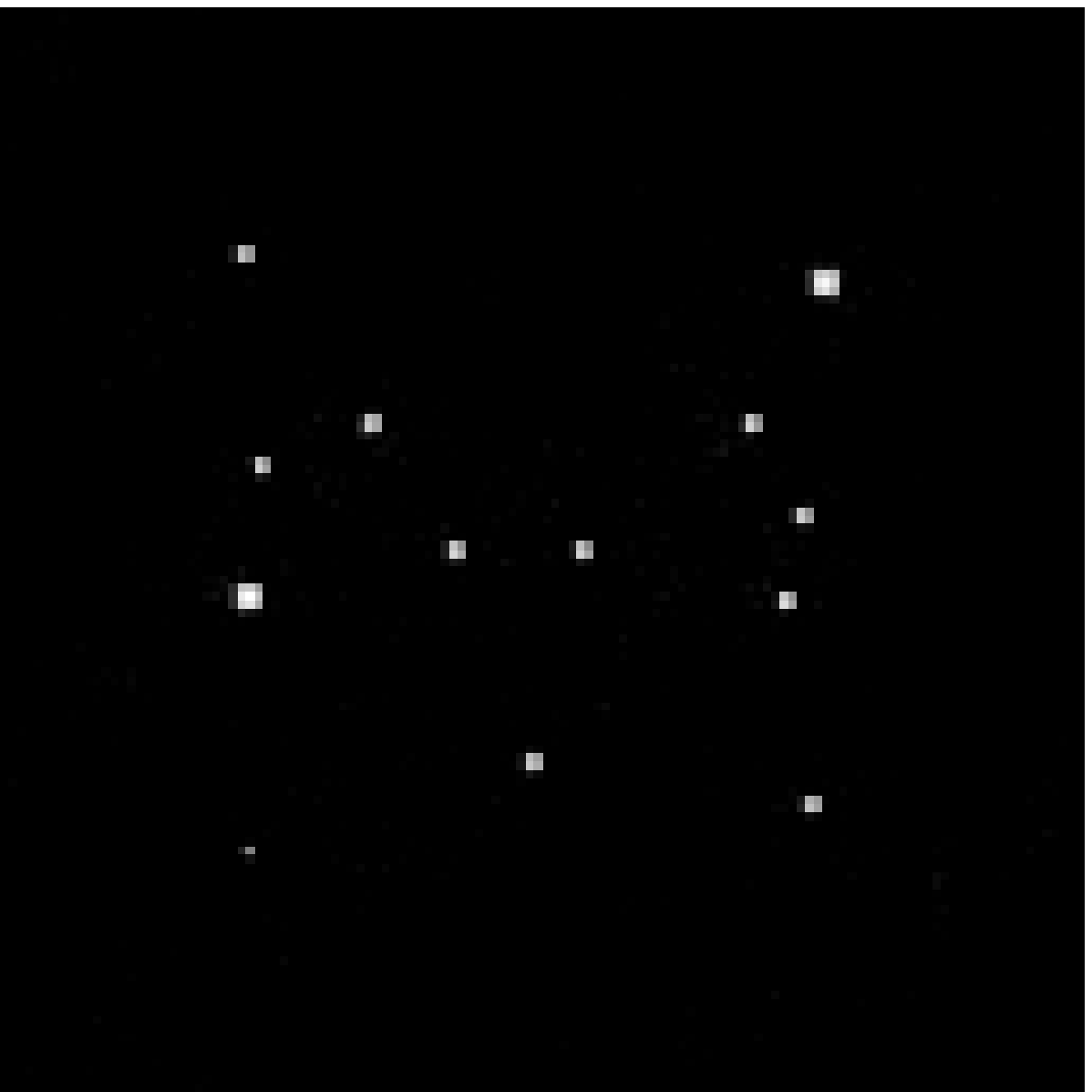}}
\end{center}
\caption{Simulation results.
\subref{sim_or}~The object image and \subref{sim_CA}~the CA.
\subref{sim_Z8_map}-\subref{sim_Z24_map}~The aberration phase map, \subref{sim_Z8_cap}-\subref{sim_Z24_cap}~the captured image, and \subref{sim_Z8_rec}-\subref{sim_Z24_rec}~the reconstructed image.
\subref{sim_Z8_map}\subref{sim_Z8_cap}\subref{sim_Z8_rec}~Simulation results with eight Zernike coefficients~($N=8$).
\subref{sim_Z16_map}\subref{sim_Z16_cap}\subref{sim_Z16_rec}~Simulation results with sixteen Zernike coefficients~($N=16$).
\subref{sim_Z24_map}\subref{sim_Z24_cap}\subref{sim_Z24_rec}~Simulation results with twenty-four Zernike coefficients~($N=24$).}
\label{sim}
\end{figure}

We demonstrated the proposed method in simulation and optical experiments.
In the simulation, we performed image reconstruction under aberrations with different strengths, as shown in Fig.~\ref{sim}.
We generated the phase map~$w$ of the aberrations by using Zernike coefficients and calculated the aberration pupil function $P$ as
\begin{align}
&w(r,\theta)=\sum_{n=1}^{N}a_n z_n(r,\theta),\\
&P ={\rm exp}(i \times w),
\label{eq_phasemap}
\end{align}
where $r$ is the radial distance, $\theta$ is the azimuthal angle, $a_n$ is the Zernike coefficients, $z_n$ denotes the Zernike series, and $N$ is the polynomial order assumed in the simulation~\cite{Schmidt2010}.
The order~$N$ was set to 8, 16, and 24 to change the strength of the aberration as in the phase maps of Figs.~\ref{sim_Z8_map}-\ref{sim_Z24_map}, where the coefficients~$a_n$ were randomly generated between $-1$ and 1.
In the simulation, the image size was set to $128\times 128$~pixels.
The object image~in Fig.~\ref{sim_or} was composed of sparsely arranged point sources with different sizes, assuming astronomical observations.
The CA in Fig.~\ref{sim_CA} was a binary random pattern.
Its transmittance ratio, which was defined as the ratio of the number of white pixels to the total pixel count on the CA, was 50~\%.
The captured images in Figs.~\ref{sim_Z8_cap}-\ref{sim_Z24_cap} were calculated by using the forward model in Eq.~(\ref{eq_forward}), where white Gaussian noise with a signal-to-noise ratio~(SNR) of 30~dB was added.
The images in Figs.~\ref{sim_Z8_rec}-\ref{sim_Z24_rec} were reconstructed from these single captured images with 100,000 iterations of the loop in Fig.~\ref{optalgorithm}.
The ambiguity of image translations in blind deconvolution was compensated for based on the cross-correlation between the object and reconstructed images.
The peak SNRs~(PSNRs) between the original and reconstructed images in Figs.~\ref{sim_Z8_rec}-\ref{sim_Z24_rec} were 36.6~dB, 42.9~dB, and 36.1~dB, respectively.
Therefore, in this simulation, the proposed method was successfully demonstrated with various strengths of aberrations.

\begin{figure}
\begin{center}
\includegraphics[scale=0.6]{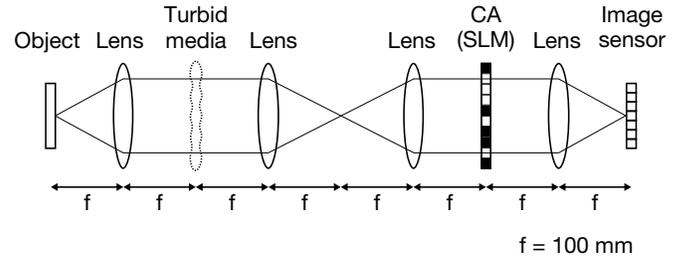}
\end{center}
\caption{Experimental setup.
SLM:~spatial light modulator.}
\label{exp_setup}
\end{figure}

\begin{figure}[t!]
\begin{center}
		\subfigure[]{\label{ex_or}\includegraphics[height=1.7cm]{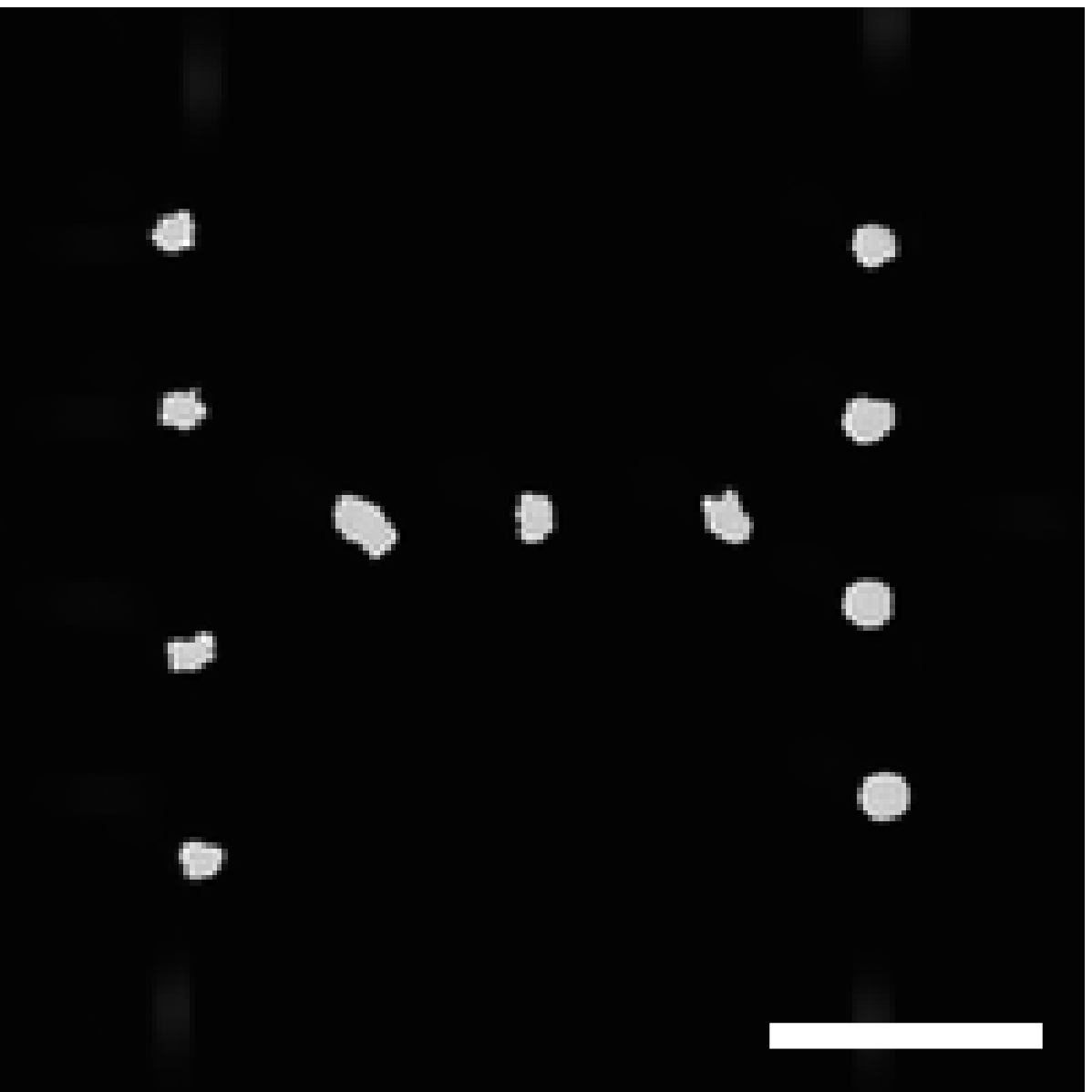}}\\
		\subfigure[]{\label{ex_CA1_cap}\includegraphics[width=1.7cm]{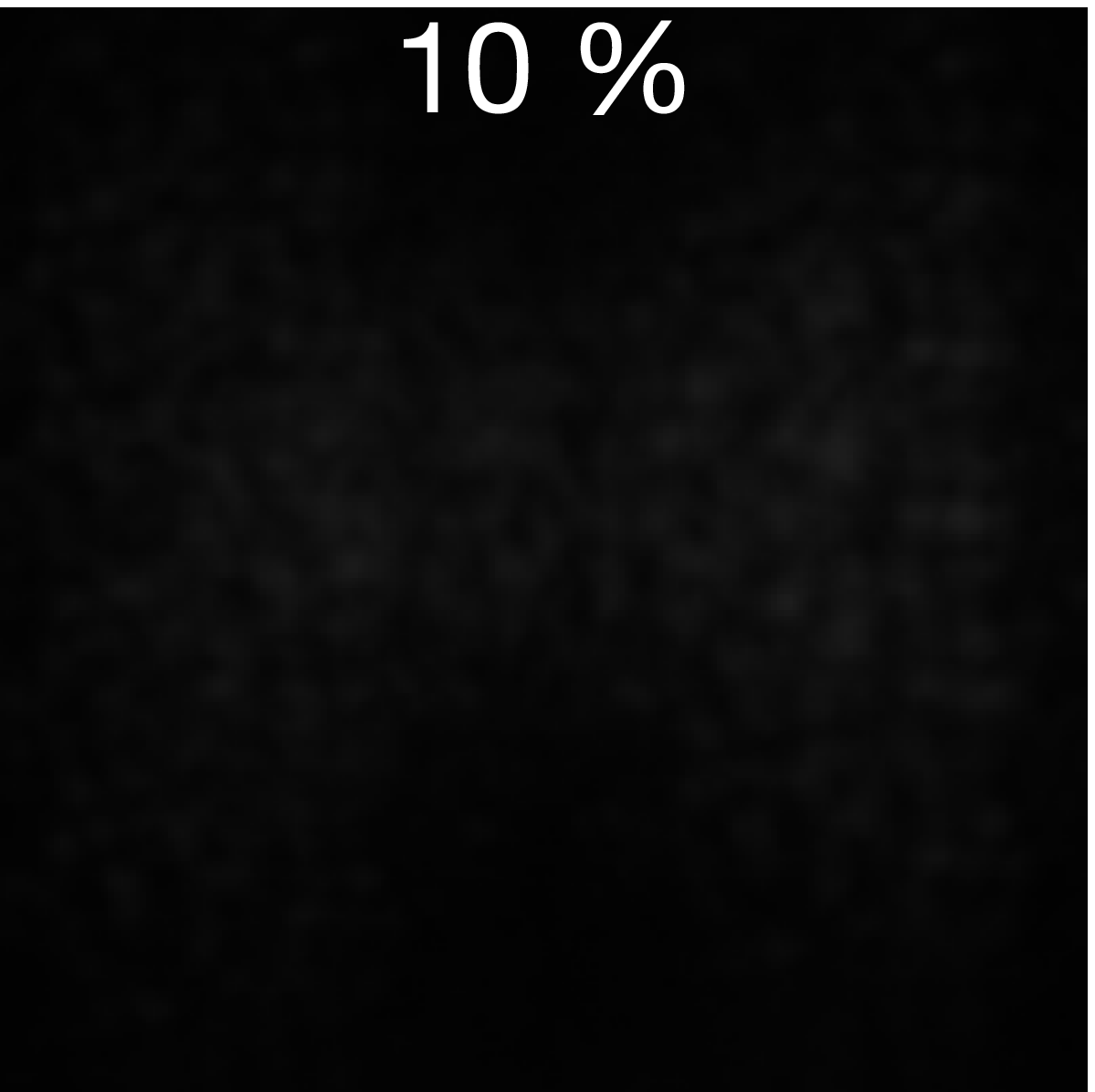}}
		\subfigure[]{\label{ex_CA2_cap}\includegraphics[width=1.7cm]{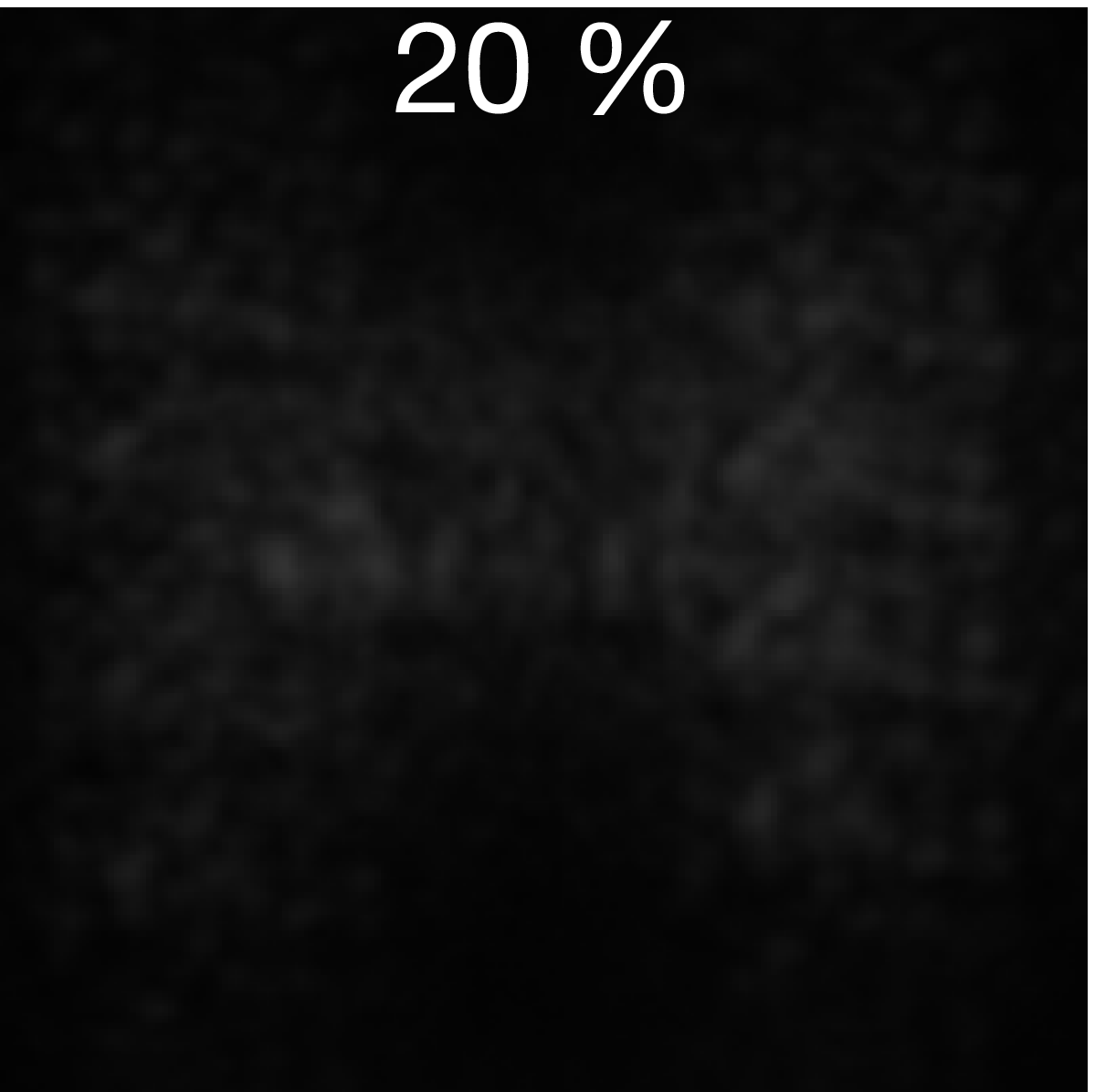}}
		\subfigure[]{\label{ex_CA3_cap}\includegraphics[width=1.7cm]{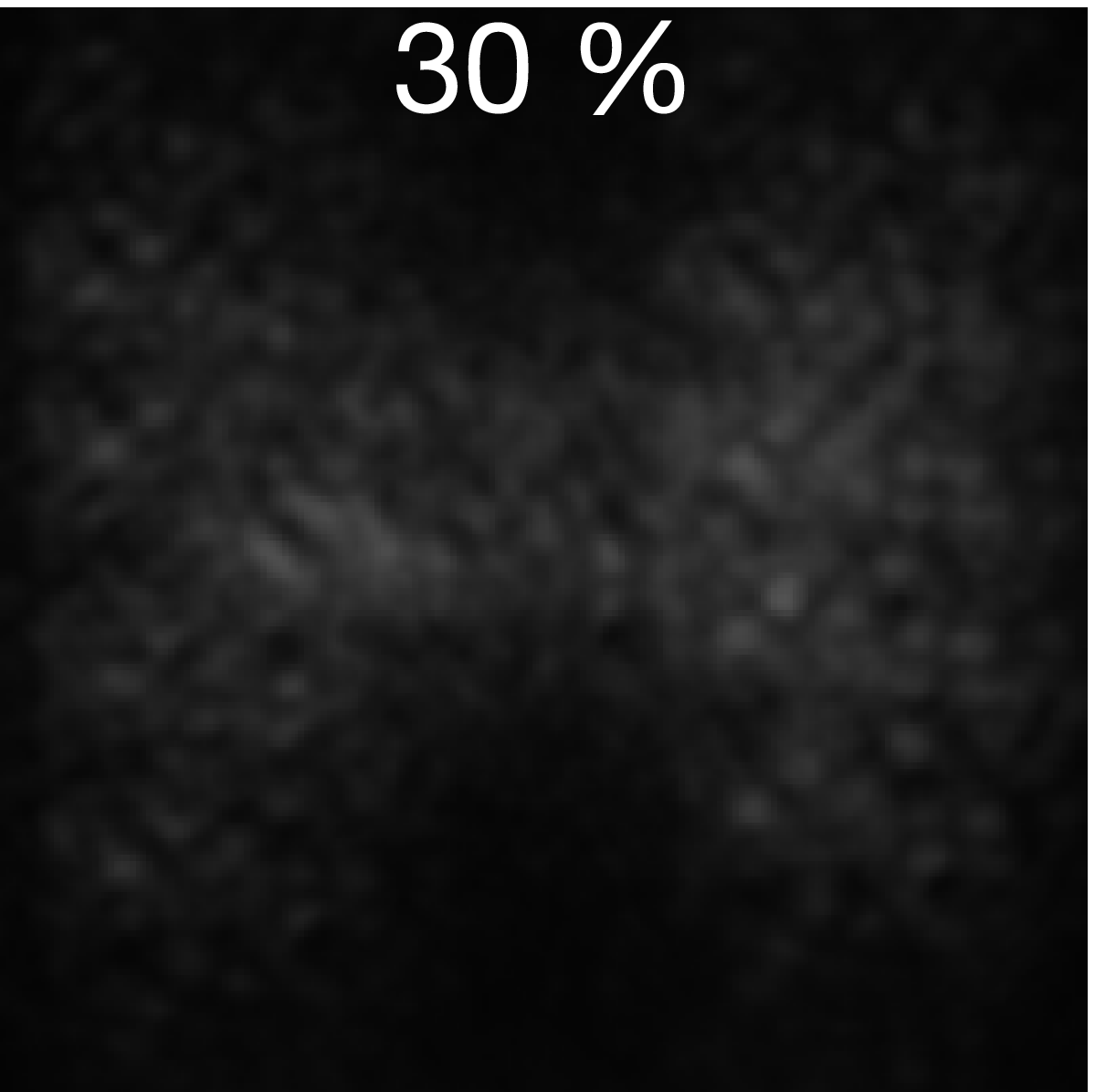}}
		\subfigure[]{\label{ex_CA4_cap}\includegraphics[width=1.7cm]{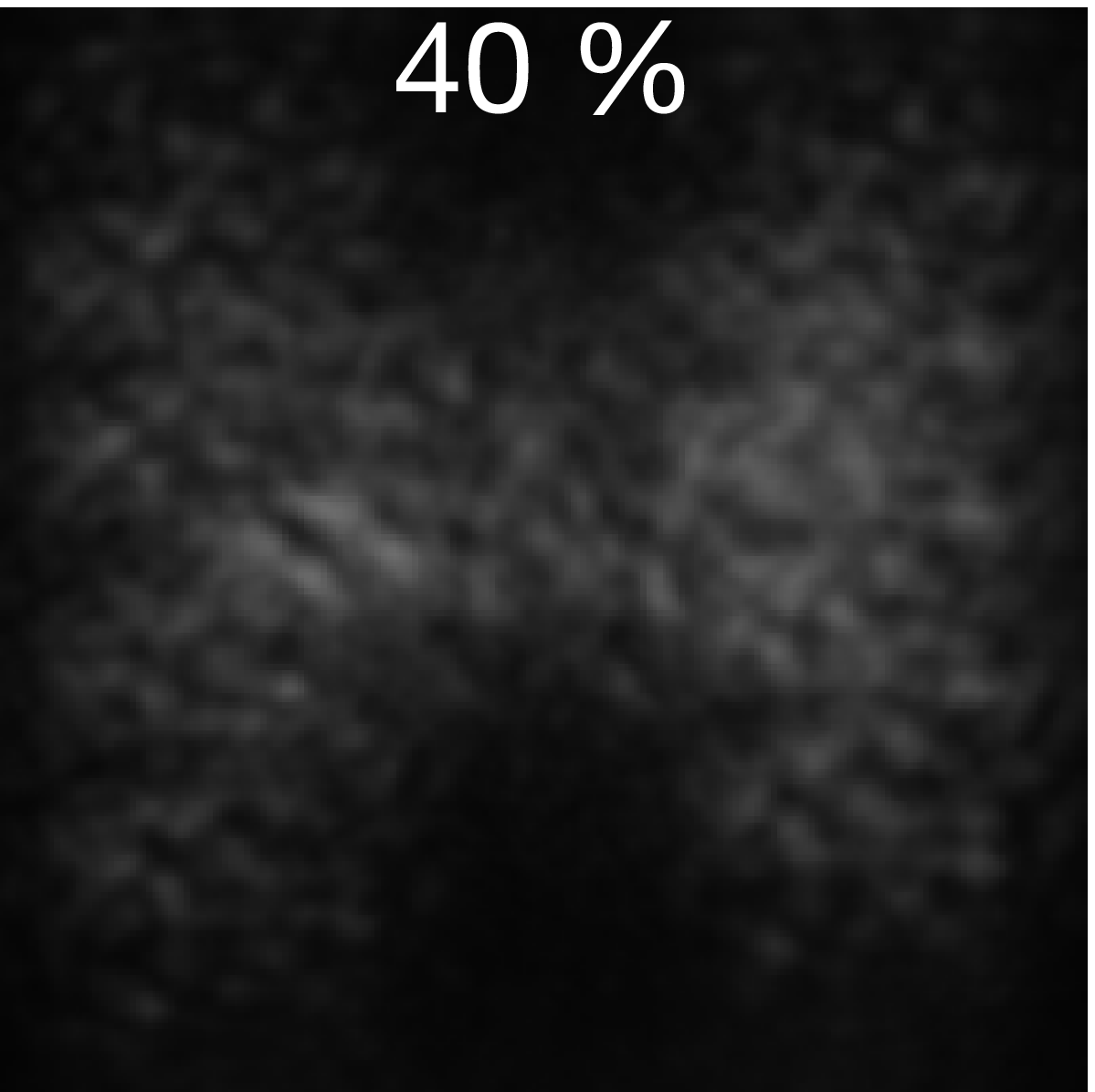}}
		\subfigure[]{\label{ex_CA5_cap}\includegraphics[width=1.7cm]{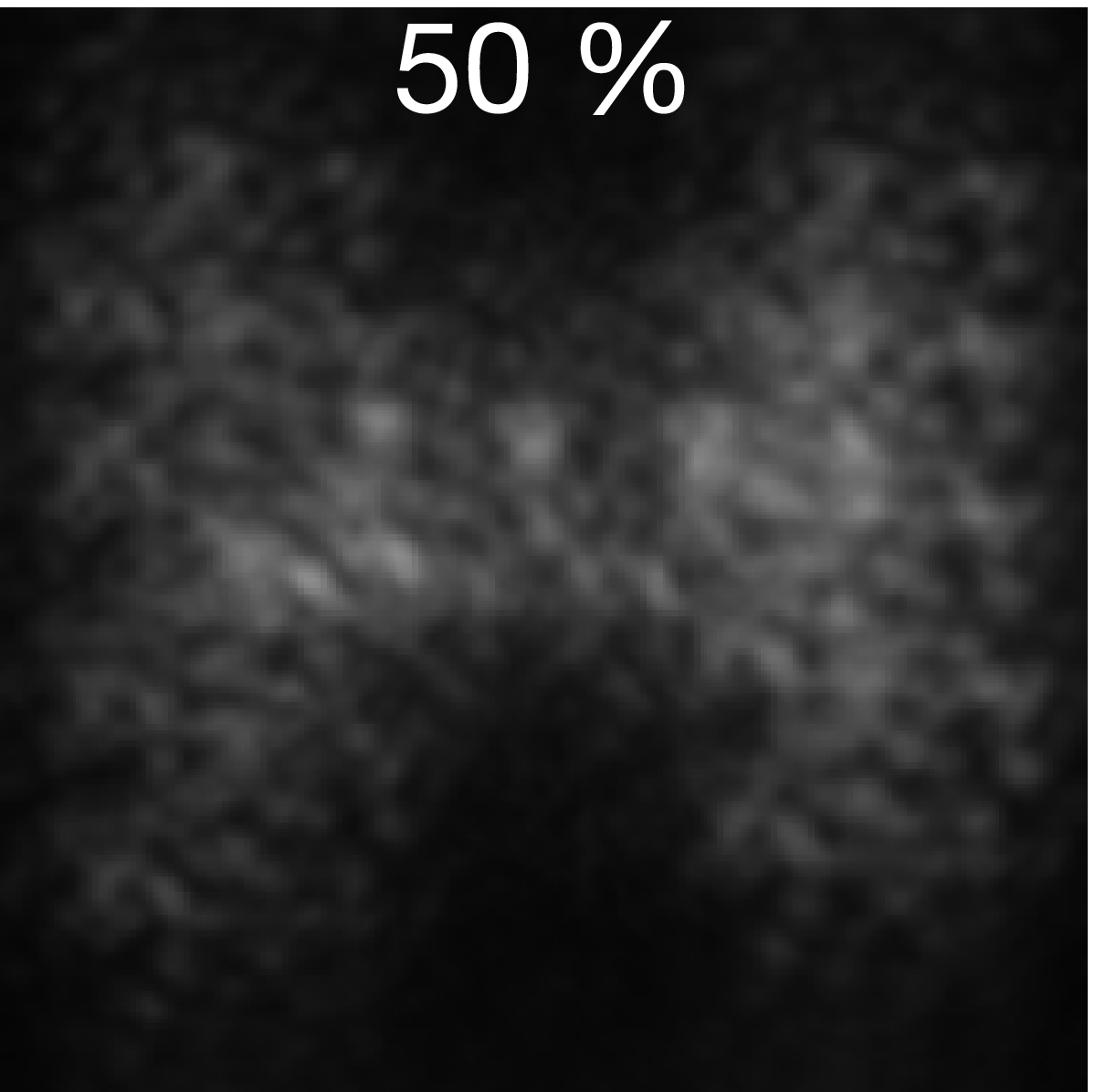}}\\
		\vspace{-0.2cm}
		\subfigure[]{\label{ex_CA6_cap}\includegraphics[width=1.7cm]{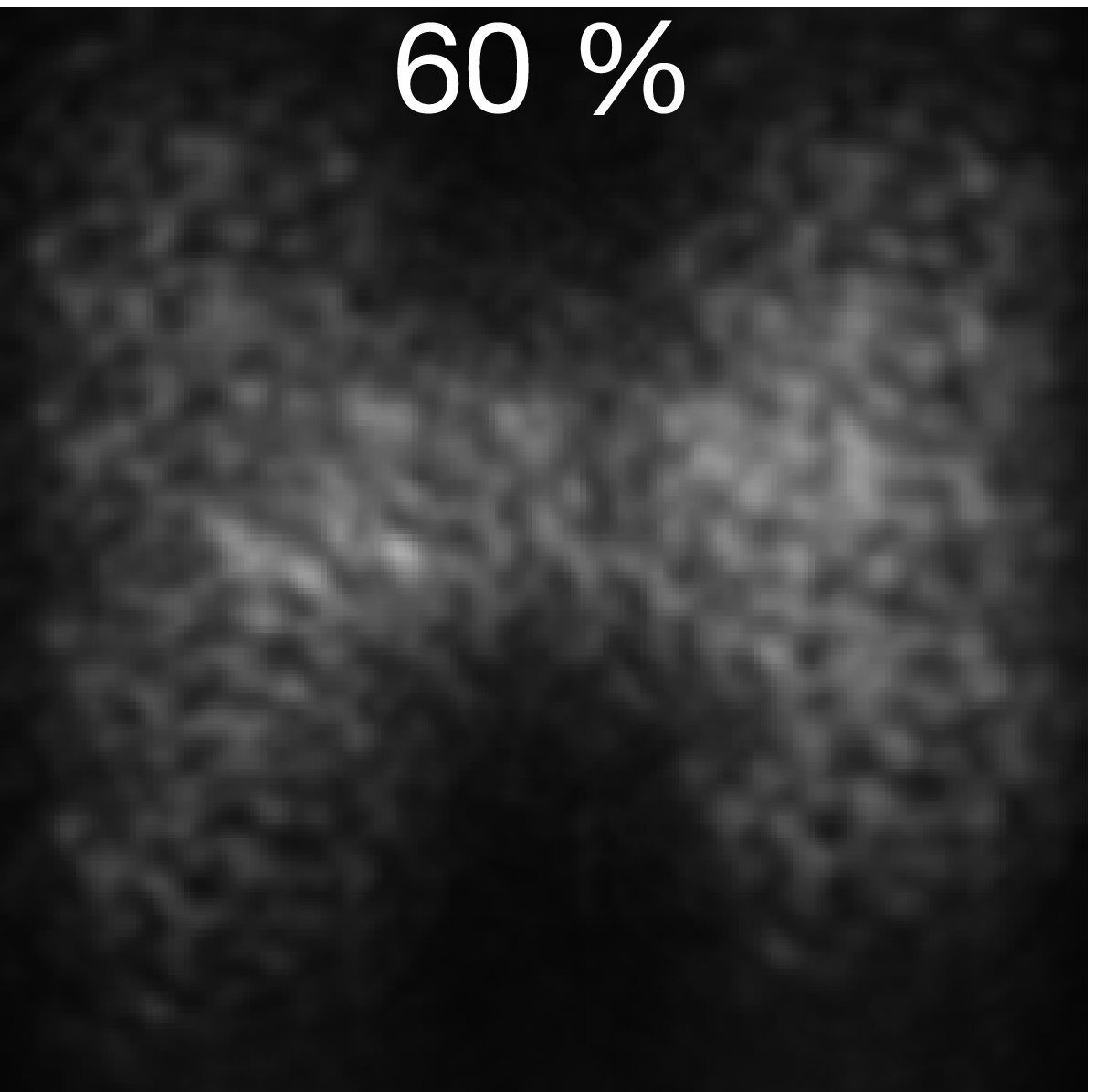}}
		\subfigure[]{\label{ex_CA7_cap}\includegraphics[width=1.7cm]{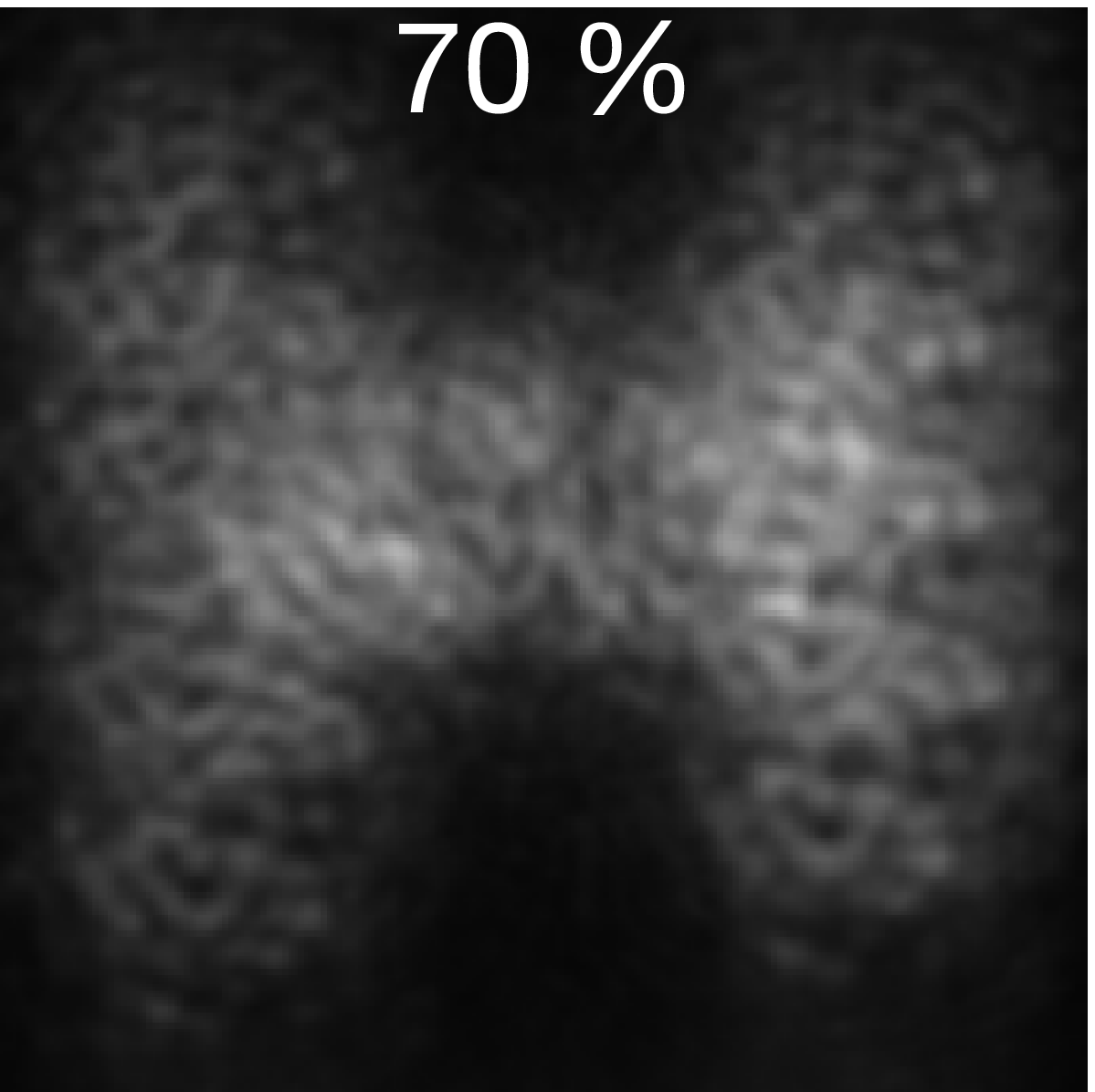}}
		\subfigure[]{\label{ex_CA8_cap}\includegraphics[width=1.7cm]{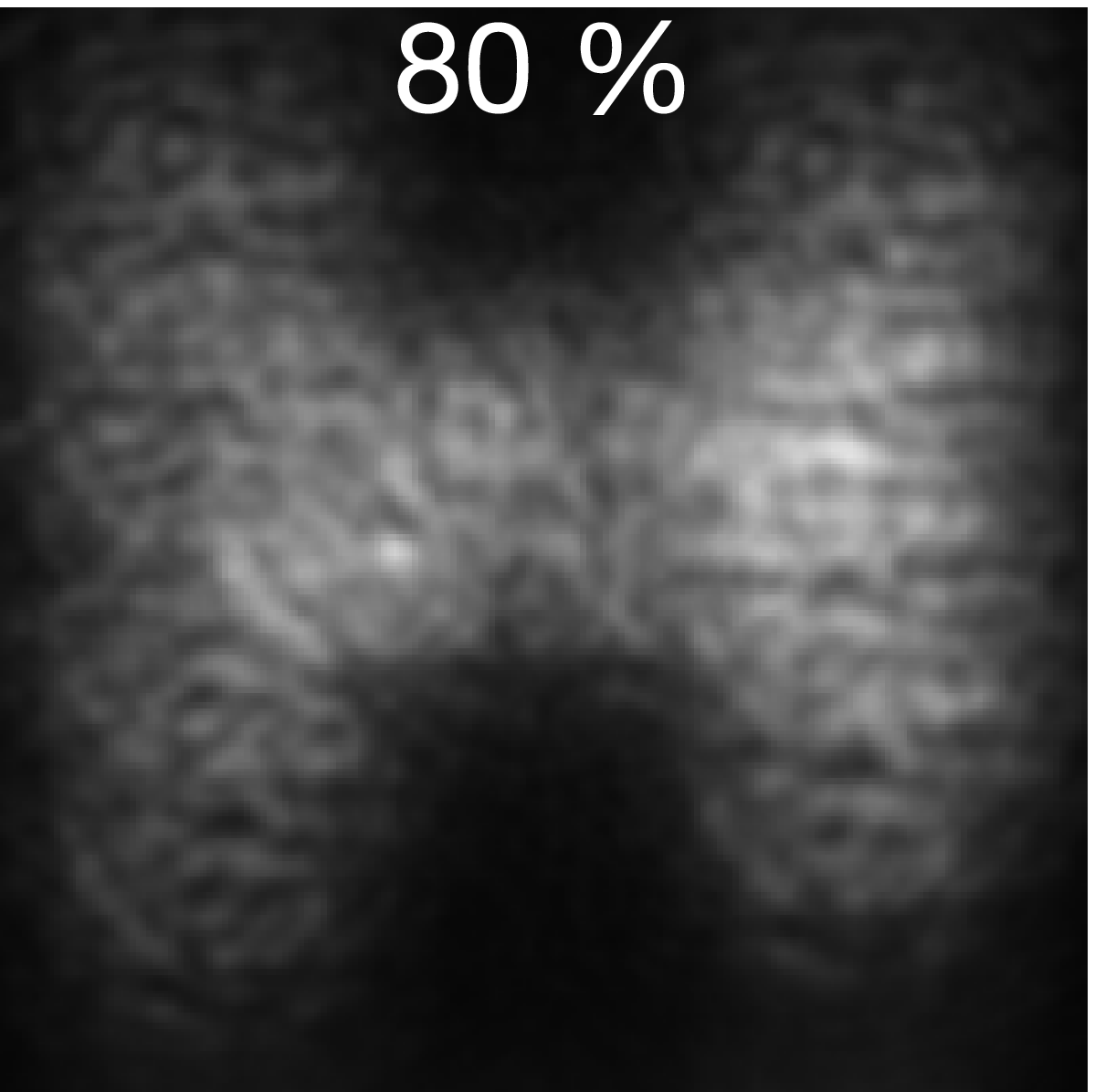}}
		\subfigure[]{\label{ex_CA9_cap}\includegraphics[width=1.7cm]{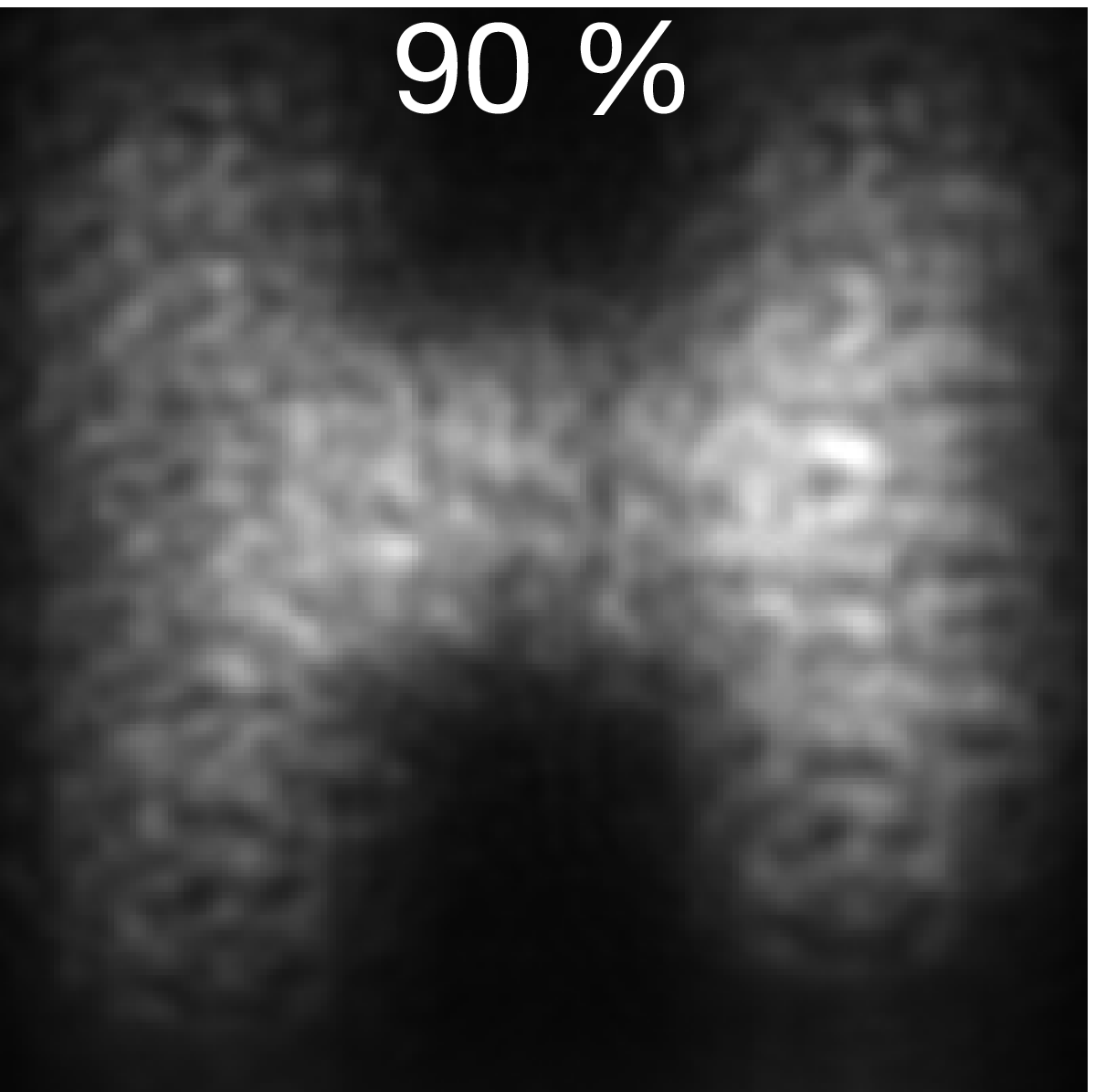}}
		\subfigure[]{\label{ex_CA10_cap}\includegraphics[width=1.7cm]{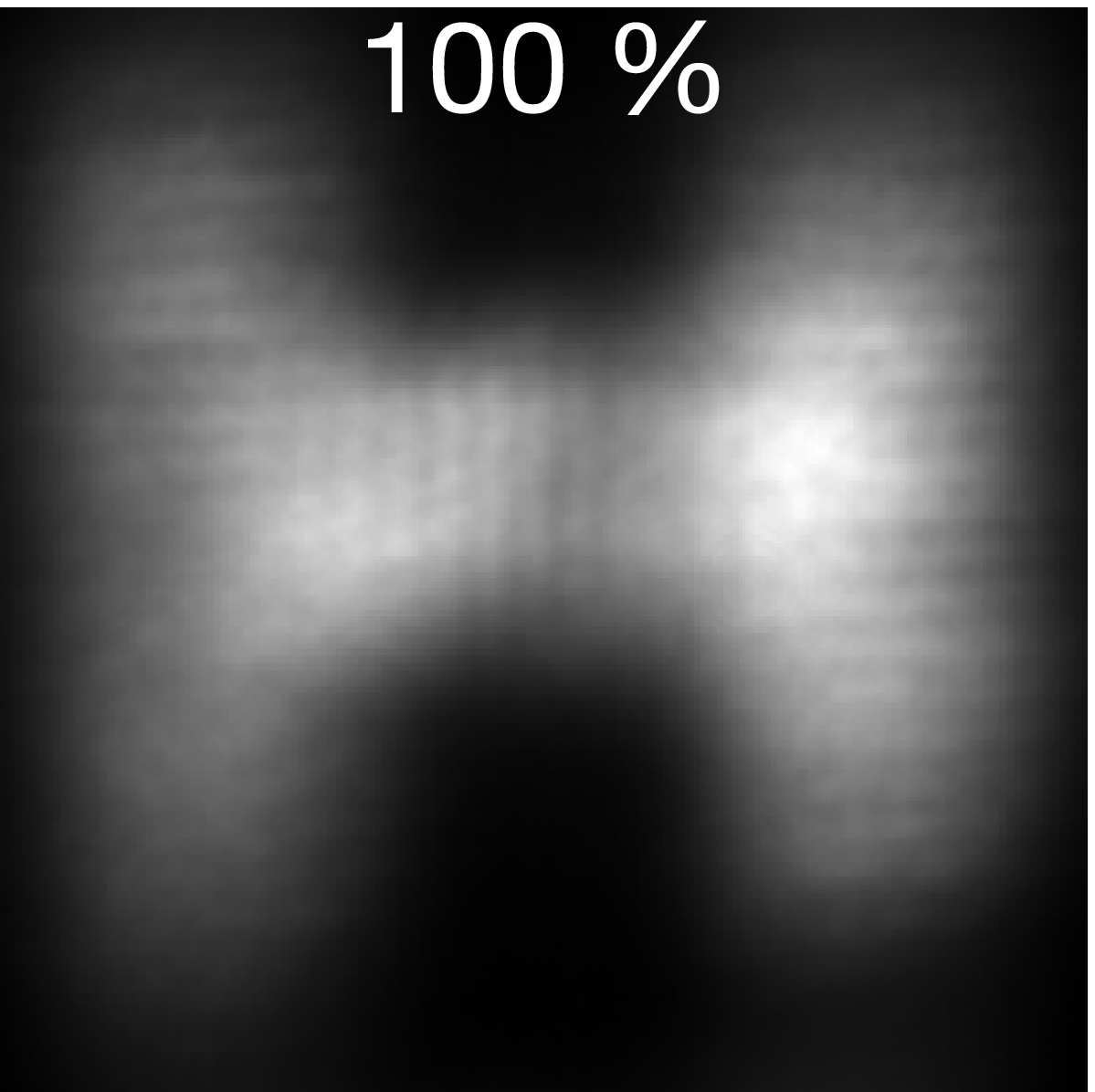}}\\
		\vspace{-0.2cm}
		\subfigure[]{\label{ex_CA1_rec}\includegraphics[height=1.7cm]{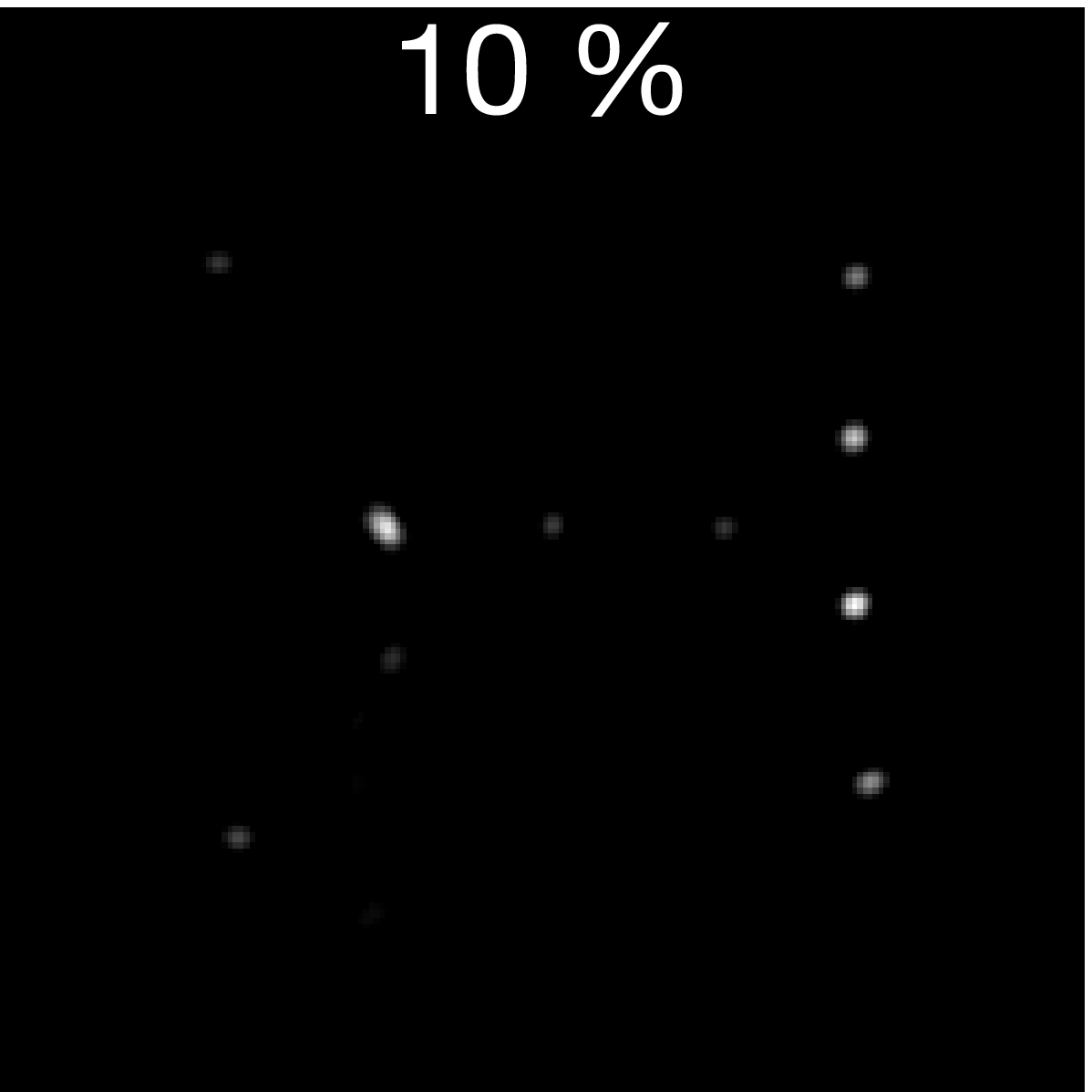}}
		\subfigure[]{\label{ex_CA2_rec}\includegraphics[height=1.7cm]{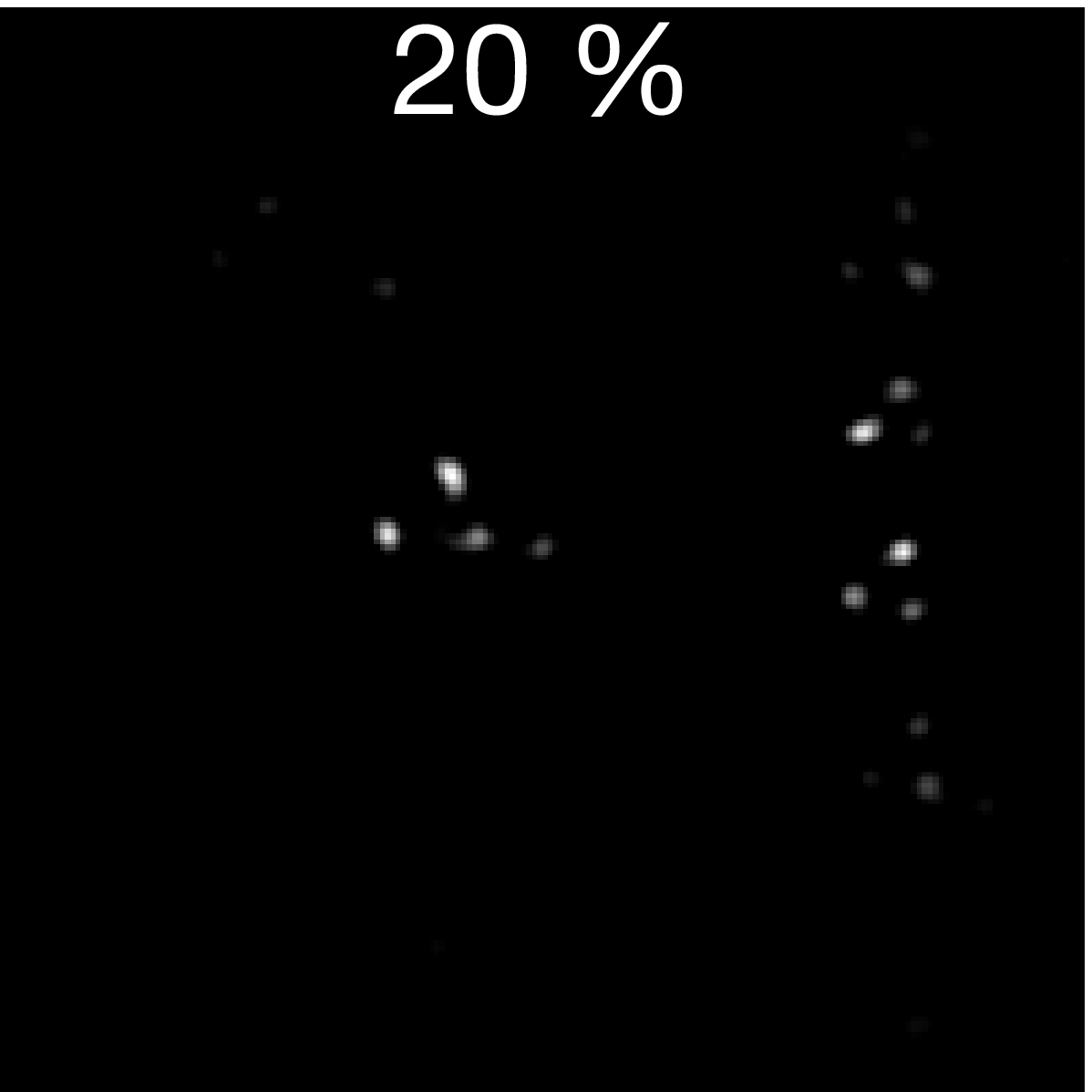}}
		\subfigure[]{\label{ex_CA3_rec}\includegraphics[height=1.7cm]{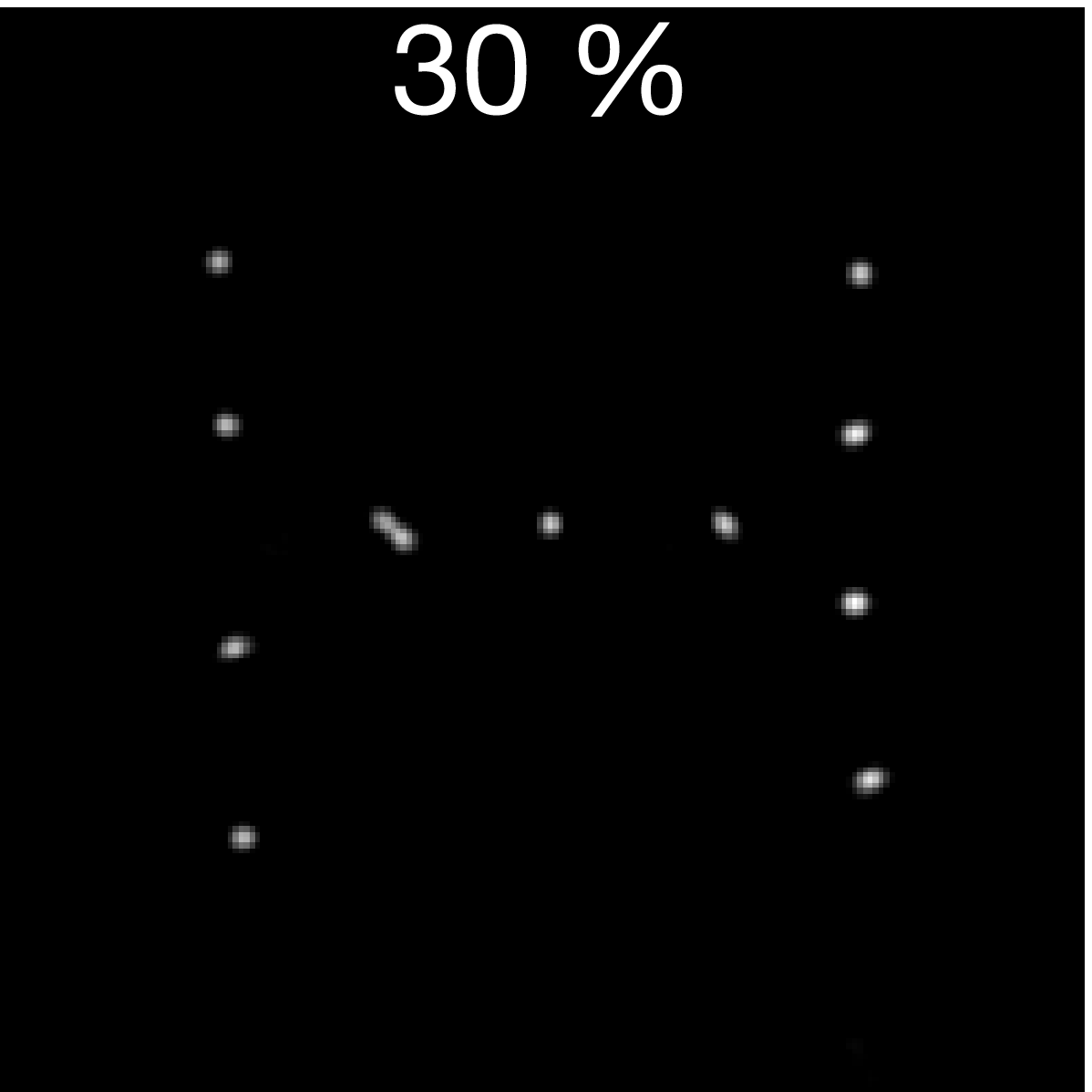}}
		\subfigure[]{\label{ex_CA4_rec}\includegraphics[height=1.7cm]{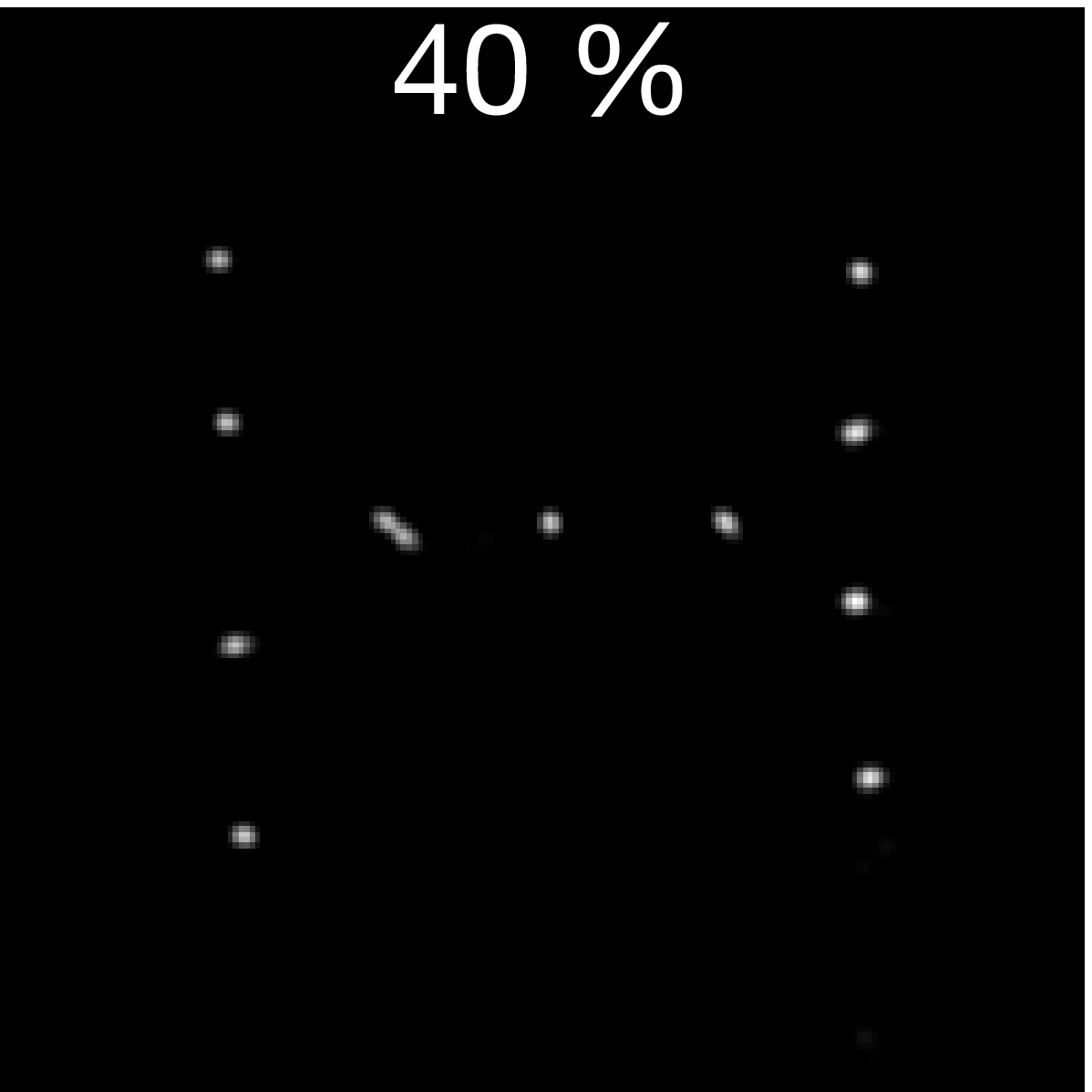}}
		\subfigure[]{\label{ex_CA5_rec}\includegraphics[height=1.7cm]{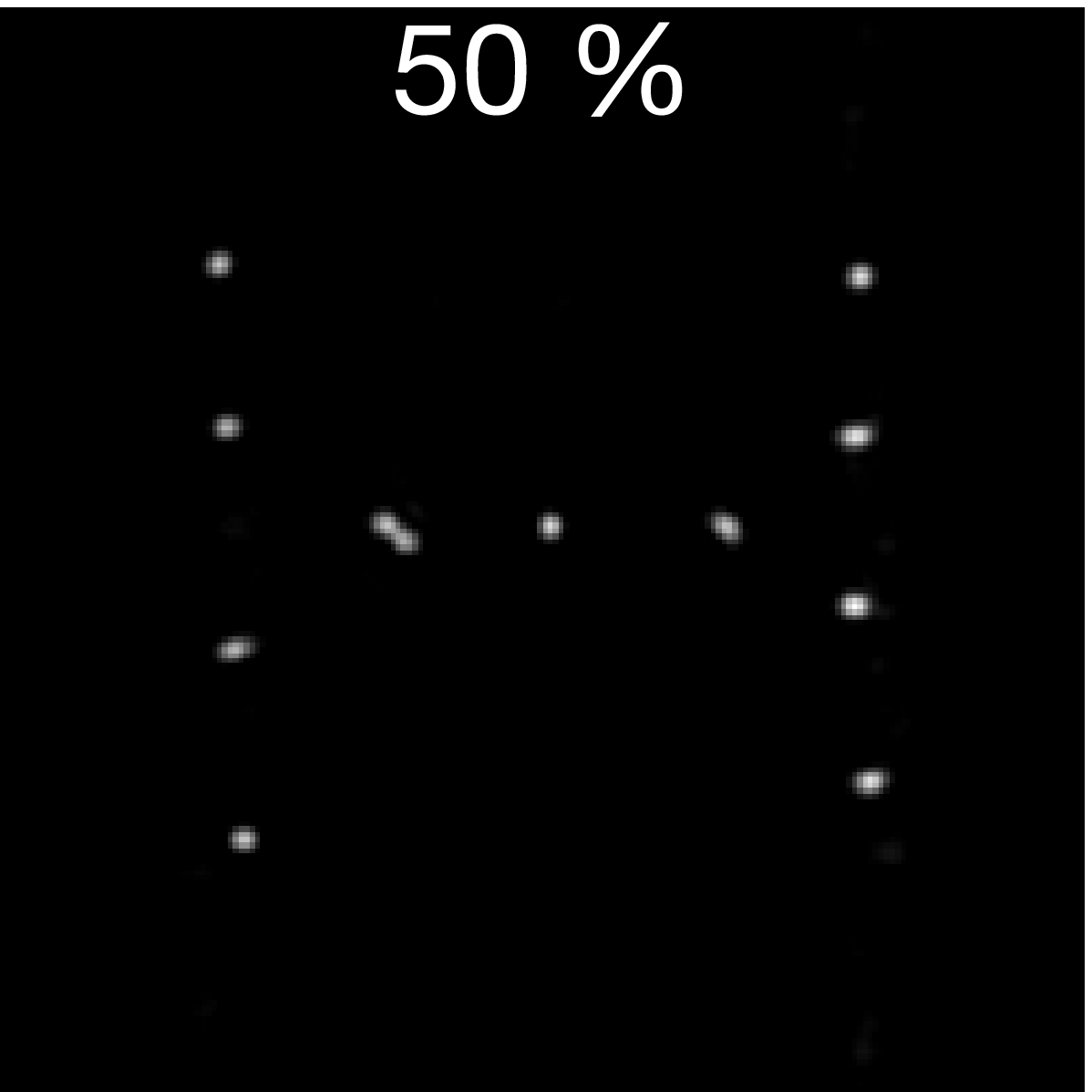}}\\
		\vspace{-0.2cm}
		\subfigure[]{\label{ex_CA6_rec}\includegraphics[height=1.7cm]{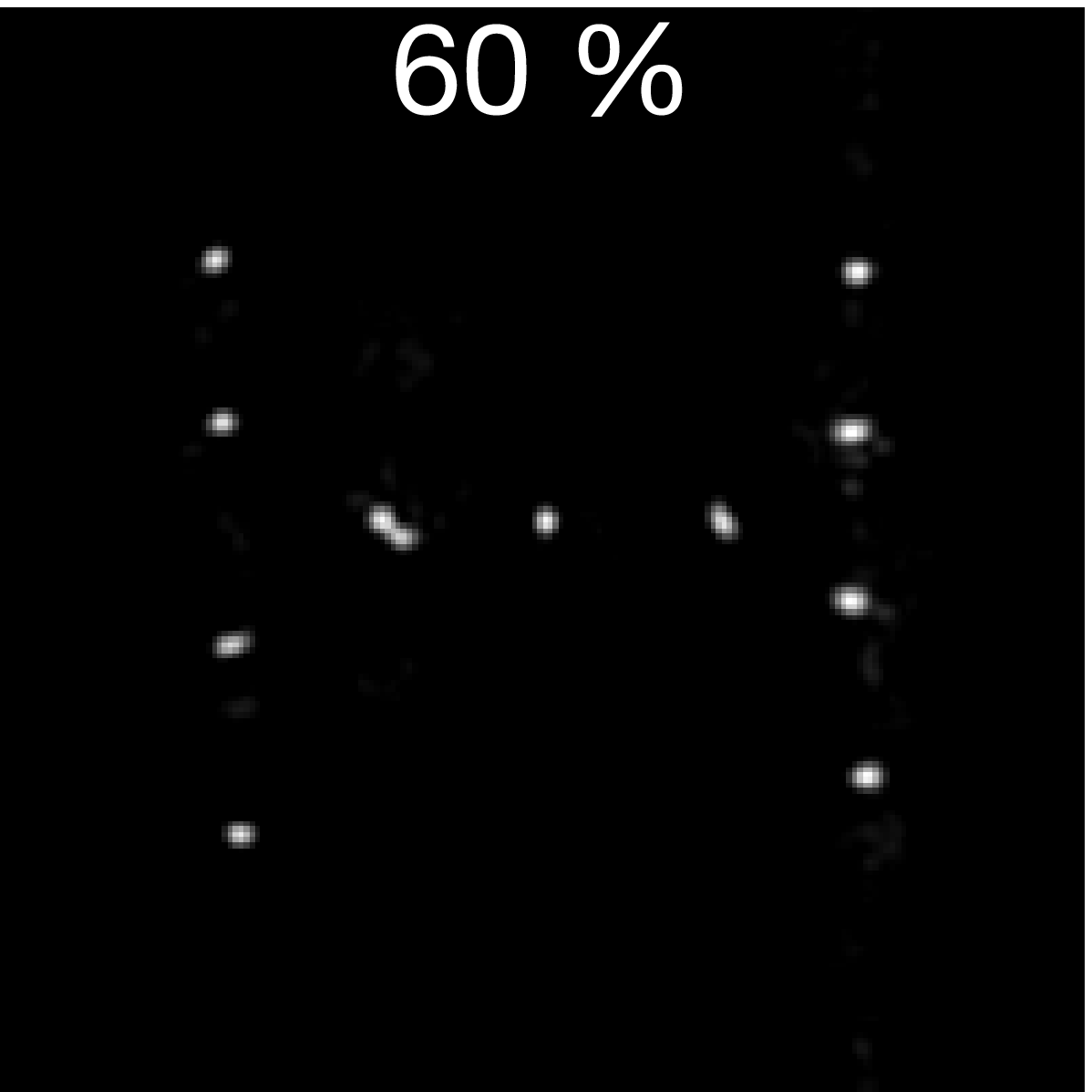}}
		\subfigure[]{\label{ex_CA7_rec}\includegraphics[height=1.7cm]{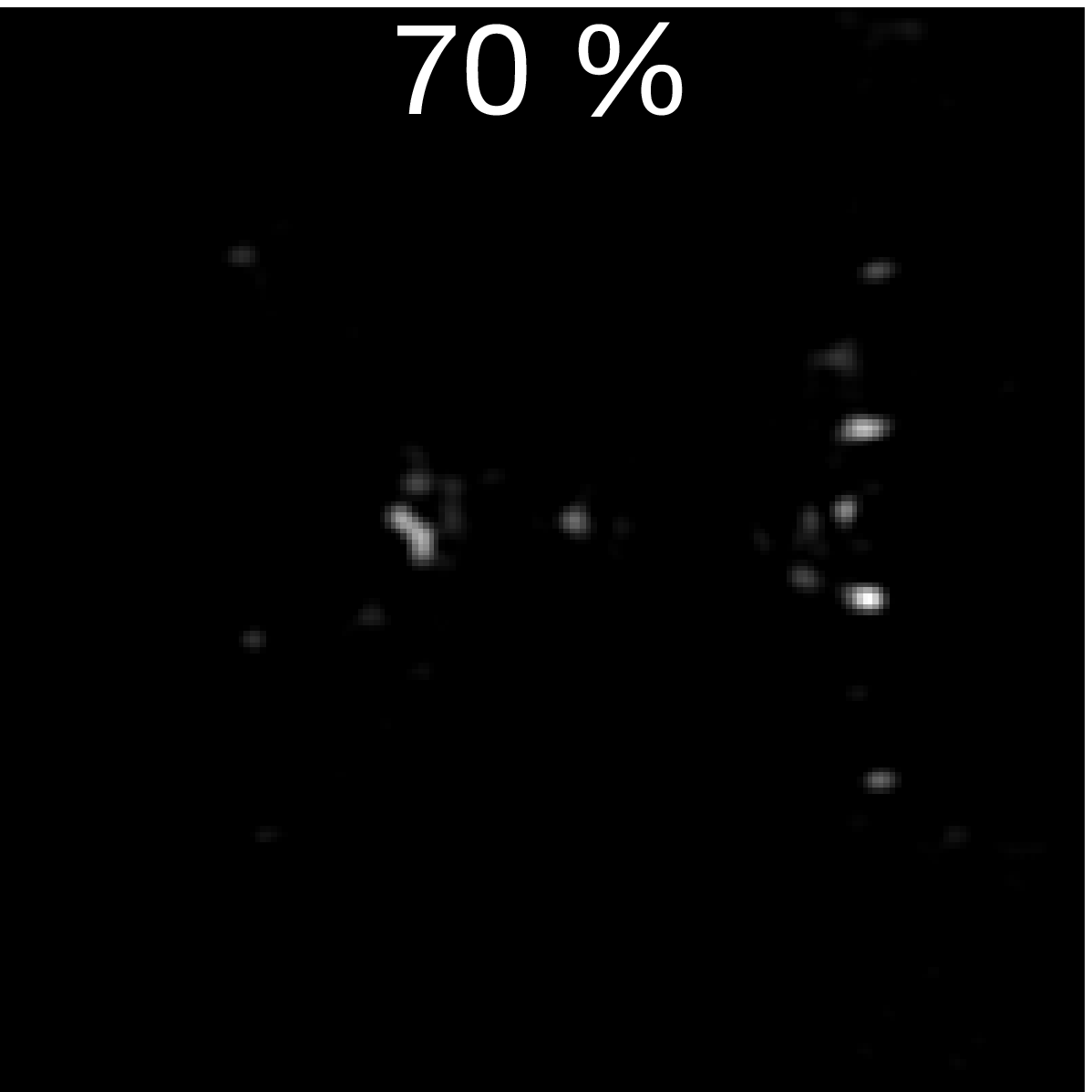}}
		\subfigure[]{\label{ex_CA8_rec}\includegraphics[height=1.7cm]{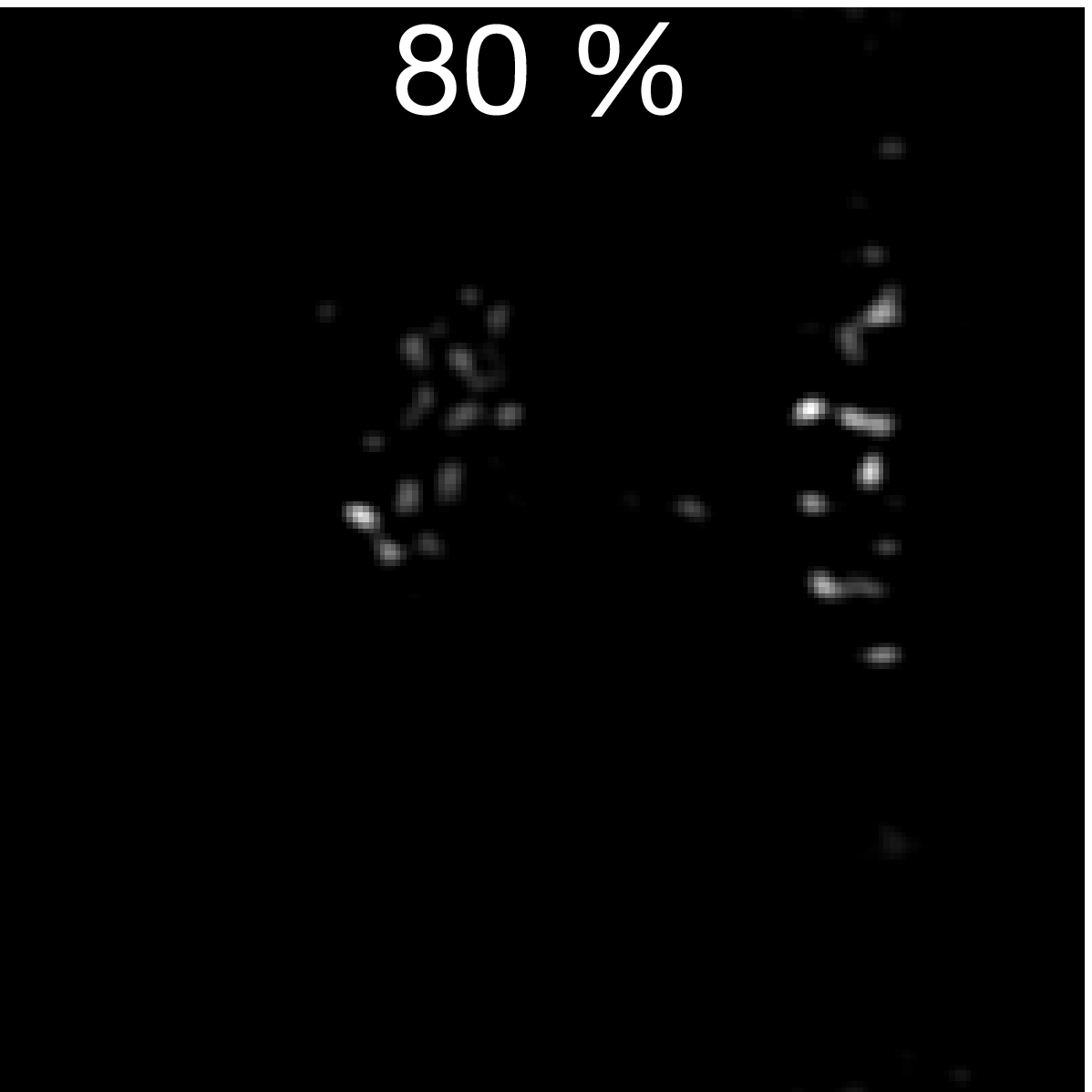}}
		\subfigure[]{\label{ex_CA9_rec}\includegraphics[height=1.7cm]{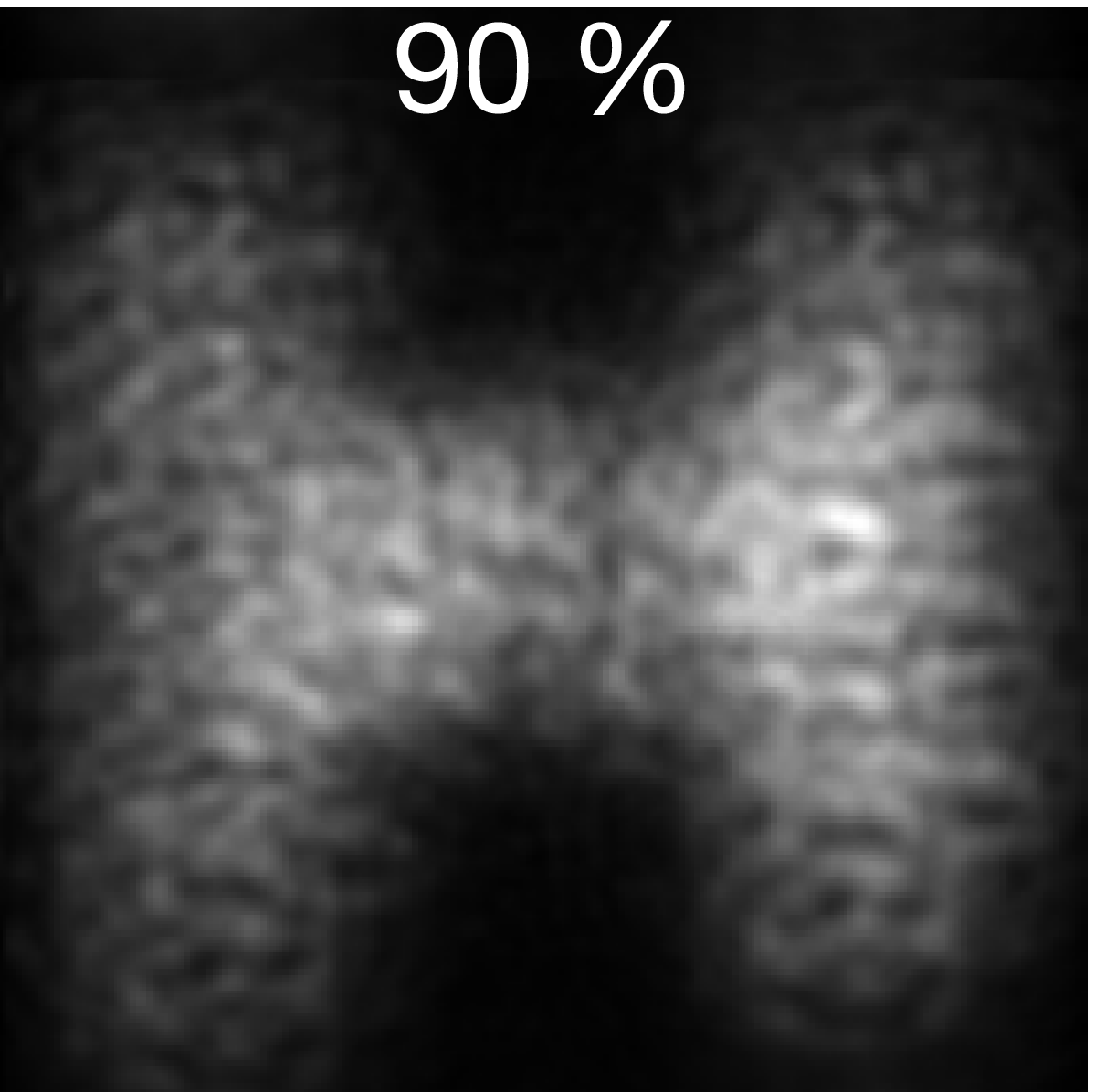}}
		\subfigure[]{\label{ex_CA10_rec}\includegraphics[height=1.7cm]{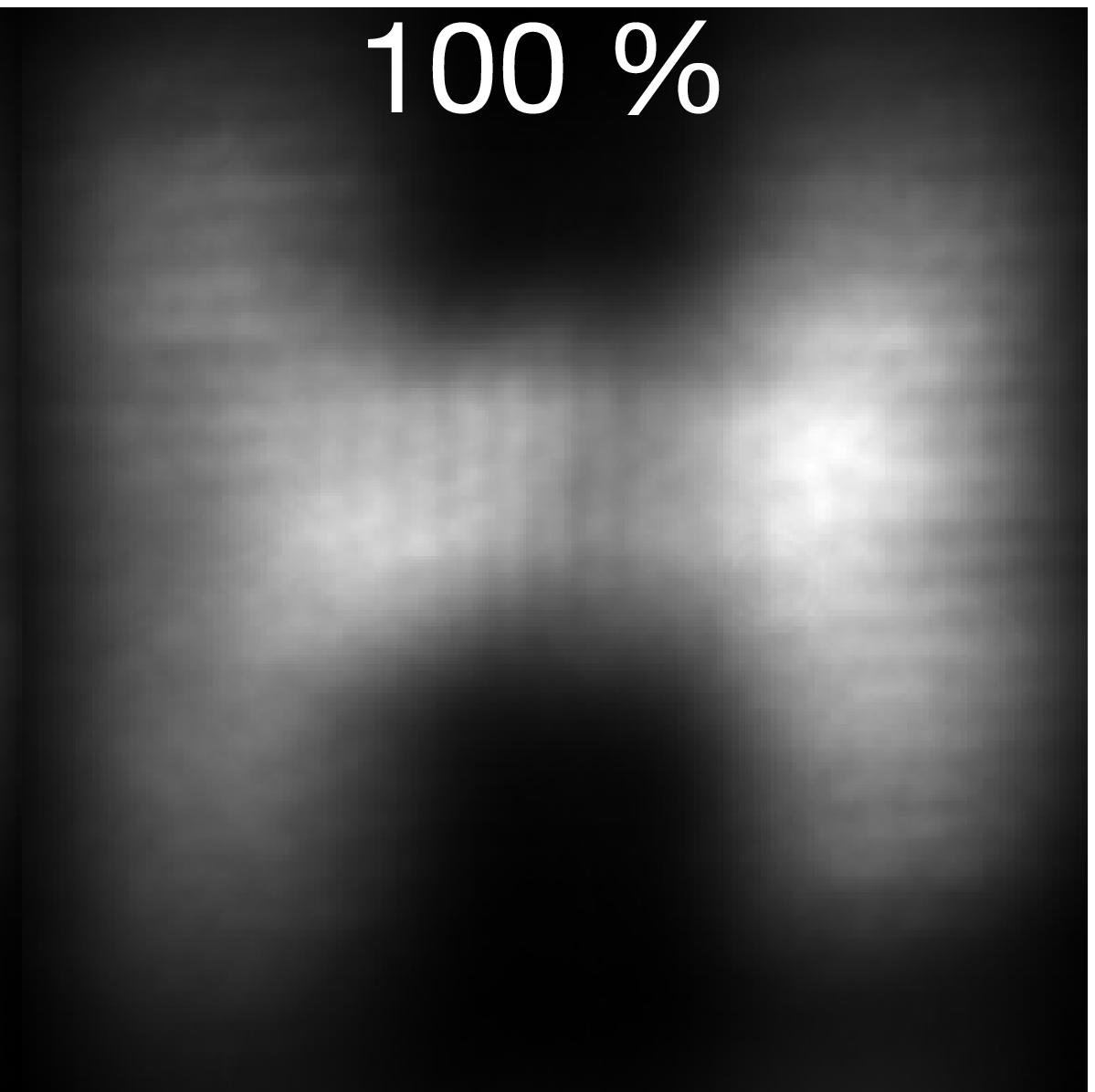}}
\end{center}
\caption{Experimental results.
\subref{ex_or}~The direct image of the object.
\subref{ex_CA1_cap}-\subref{ex_CA10_cap}~The captured images with different transmittance ratios.
\subref{ex_CA1_rec}-\subref{ex_CA10_rec}~The reconstructed images from Figs.~\subref{ex_CA1_cap}-\subref{ex_CA10_cap}.
The transmittance ratios are shown in the upper parts on each of the images.
The length of the scale bar on Fig.~\subref{ex_or} is 0.5~mm.}
\label{exopt}
\end{figure}

Next, we experimentally demonstrated the proposed method with the optical setup shown in Fig.~\ref{exp_setup}.
The setup consisted of two 4f systems, where the upstream one was for introducing the aberrations and the downstream one implemented imaging through the CA.
The focal length of lenses for the 4f systems was 100~mm.
The object was composed of multiple holes on a flat aluminum foil, as shown in Fig.~\ref{ex_or}, and it was illuminated with spatially incoherent light from a green light emitting diode~(LED:~LST1-01F06-GRN1-00 manufactured by New Energy, wavelength:~530~nm).
Light from the object was disturbed through the upstream 4f system, where a lens with a focal length of 200~mm was located on the pupil plane to serve as turbid media in Fig.~\ref{exp_setup} for the shift-invariant aberration.
The aberrated image was captured through the downstream 4f system with the CA implemented by a spatial light modulator~(SLM:~LC2012 manufactured by HOLOEYE, pixel count:~$1024\times768$, pixel pitch:~36~\textmu m) in the amplitude mode and a monochrome image sensor~(PL-D7512 manufactured by PixeLink, pixel count:~$4096\times 3000$, pixel pitch:~3.45~\textmu m).

In the experiment, binary random patterns with different transmittance ratios were used for the CA.
As shown in Fig.~\ref{exopt}, the transmittance ratio was changed from 10~\% to 100~\% with intervals of 10~\%.
The exposure time was set to 0.13~seconds at all the transmittance ratios, and it was chosen for the transmittance ratio of 100~\%~(the brightest case) to prevent light saturation.
The captured images with each transmittance ratio are shown in Figs.~\ref{ex_CA1_cap}-\ref{ex_CA10_cap}, where the point sources on the object plane were not visually recognizable.
These captured images were the results obtained after downsampling with a factor of three and cropping the central area of $200\times 200$~pixels.  
The reconstruction results from the captured images are shown in Figs.~\ref{ex_CA1_rec}-\ref{ex_CA10_rec}.
By comparisons between them and Fig.~\ref{ex_or}, the object was successfully reconstructed based on the proposed method with the transmittance ratios between 30~\% and 60~\%.
Possible reasons for the failure cases may be noisy measurements due to lower light intensities when the transmittance ratios were less than 30~\%, and insufficient support constraints with denser CAs when the transmittance ratios were larger than 60~\%, respectively.

\section{Conclusion}

In summary, we proposed and demonstrated a single-shot imaging method for blind deconvolution with a CA.
Single-shot blind deconvolution is an ill-posed inverse problem.
In the proposed method, the CA located on the pupil plane in the imaging optics works as the support constraint in the inverse problem.
An object is captured with the CA-based imaging optics through turbulence, and it is reconstructed from the single captured image by using the PIE-based algorithm incorporating the CA-based support constraint.
The proposed method was numerically and experimentally demonstrated by reconstructing point sources from images captured through severe aberrations.
Our method realized single-shot blind deconvolution by using compact and low-cost imaging hardware with a high light efficiency.

The reconstruction algorithm in this study assumed shift-invariant and adsorption-free turbulence.
These assumptions may be resolved by modifying the algorithm although they are acceptable in various applications.
The demonstrations here can be directly applied to astronomical observations and biomedical fluorescence microscopy, where many imaging targets are composed of incoherent point sources~\cite{Bobin2008, Bertero2009, Betzig2006, Rust2006, Tehrani2015}.
Therefore, our single-shot blind deconvolution method will contribute to various fields, including astronomy and biology.

\end{document}